\documentclass[aps,prd,twocolumn,superscriptaddress,showpacs]{revtex4}

\usepackage{graphicx}
\usepackage{times}
\usepackage{type1cm}
\usepackage{eso-pic}
\usepackage{color}

\newcommand{\be}{\begin{equation}}
\newcommand{\ee}{\end{equation}}
\newcommand{\bea}{\begin{eqnarray}}
\newcommand{\eea}{\end{eqnarray}}

\newcommand{\tQ}{\widetilde{Q}}

\newcommand{\bq}{\mathbf{q}}
\newcommand{\bk}{\mathbf{k}}
\newcommand{\bx}{\mathbf{x}}

\begin{document}

\title{Calibrating the Baryon Oscillation Ruler for Matter and Halos}

\author{Nikhil Padmanabhan}
\email{NPadmanabhan@lbl.gov}
\affiliation{Physics Division, Lawrence Berkeley National Laboratory, 1 Cyclotron Rd., Berkeley, CA 94720}

\author{Martin White}
\email{mwhite@berkeley.edu}
\affiliation{Departments of Physics and Astronomy, 601 Campbell Hall,
University of California Berkeley, CA 94720\\
Physics Division, Lawrence Berkeley National Laboratory, 1 Cyclotron Rd.,
Berkeley, CA 94720}

\date{\today}

\begin{abstract}
We characterize the nonlinear evolution of the baryon acoustic feature
as traced by the dark matter and halos, using a combination 
of perturbation theory and N-body simulations. We confirm that the acoustic peak
traced by the dark matter 
is both broadened and shifted as structure forms, and that this shift 
is well described by second-order perturbation theory. 
These shifts persist for dark matter halos, and are
a simple function of halo bias, with the shift (mostly) increasing with 
increasing bias. 
Extending our perturbation theory results to halos with 
simple two parameter bias models (both in Lagrangian and Eulerian space) quantitatively 
explains the observed shifts. In particular, we demonstrate that there are
additional terms that contribute to the shift that are absent for 
the matter. At $z=0$ for currently favored cosmologies, the matter shows
shifts of $\sim 0.5\%$,  $b=1$ halos
shift the acoustic scale by $\sim 0.2\%$, while  $b=2$ halos shift it by $\sim 0.5\%$;
these shifts decrease by the square of the growth factor $D(z)$ at higher redshifts.
These results are easily generalized to galaxies within the halo
model, where we show that simple galaxy models show marginally larger shifts than 
the correspondingly biased halos, due to the contribution of satellites in high mass halos.
While our focus here is on real space, our results make specific predictions
for redshift space.  For currently favored cosmological models, we find that
the shifts for halos at $z=0$ increase by $\sim 0.3\%$; at high $z$, they
increase by $\sim 0.5\%\ D^2$.
Our results demonstrate that these theoretical systematics are smaller
than the statistical precision of upcoming surveys, even if one ignored the corrections
discussed here. Simple modeling, along the lines discussed here, has the potential 
to reduce these systematics to below the levels of cosmic variance limited surveys.
    
\end{abstract}


\maketitle

\section{Introduction}

It has been known for many years that the coupling of photons and baryons
in the early universe results in an almost harmonic series of oscillations
in the matter power spectrum \cite{PeeYu70,SunZel70,DorZelSun78} with a
scale set by the sound horizon, $s\sim100\,$Mpc
(see \cite{EisHu98,MeiWhiPea99} for a detailed description of the physics in
 modern cosmologies and \cite{ESW07} for a comparison of Fourier and
 configuration space pictures).
This feature can be used as a `standard ruler' to measure the expansion rate
of the Universe, and this baryon acoustic oscillation (BAO) method
is an integral part of current and next-generation dark energy experiments.

While the early Universe physics is linear and well understood, the low
redshift observations are complicated by the nonlinear evolution of 
matter and the non-trivial relation between galaxies and dark matter.
The nonlinear evolution leads to a damping of the oscillations on small
scales \cite{Bha96a,Bha96b,MeiWhiPea99,ESW07,CroSco08,Mat08,SSS08}
and the generation of a small out-of-phase component
\cite{ESW07,CroSco08,Mat08,SSS08,Seo08,PadWhiCoh09}.
The damping of the linear power spectrum (or equivalently the smoothing of
the correlation function) reduces the contrast of the feature and thereby
the precision with which the size of ruler may be measured.
The out-of-phase component corresponds to a shift in the acoustic scale
which would bias the distance measure if it were not taken into account.

In this paper we investigate the behavior of the acoustic signal in
biased tracers of the nonlinear mass field.
We find that biased tracers have different shifts than the matter, and
discuss how these shifts can be modeled and significantly reduced.

\section{Preliminaries}

\begin{figure}
\begin{center}
\resizebox{3in}{!}{\includegraphics{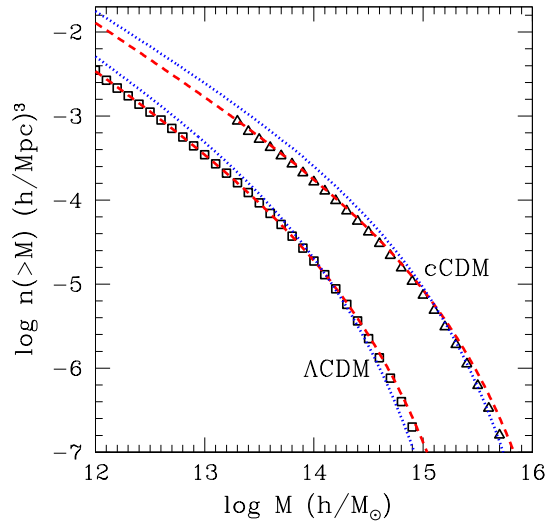}}
\end{center}
\vspace{-0.2in}
\caption{The mass functions at $z=0$ for our $c$CDM  (triangles) and
$\Lambda$CDM (squares) cosmologies, along
with two commonly used fitting functions due to Press \& Schechter
\protect\cite{PS} (dotted, blue) and Sheth \& Tormen \protect\cite{STbias}
(dashed, red).  Since we are using the sum of the particles in a FoF group
for our definition of mass, we do not expect perfect agreement with either
fitting function. The simulations for the $\Lambda$CDM cosmology are discussed
in \S\ref{sec:implications}.}
\label{fig:massfn}
\end{figure}

One of the challenges of calibrating systematic effects in BAO is the very
large scale of the acoustic feature, which requires huge volumes to be
surveyed/simulated and reduces the effects of nonlinearities and astrophysical
uncertainties.
While the latter is what makes BAO an attractive standard ruler, the
combination makes it challenging to measure systematics with any statistical
precision.  To avoid these issues, we start with a toy cosmology ($c$CDM,
see also \cite{CarWhiPad09}) which has
$\Omega_M=1$, $\Omega_B=0.4$, $h=0.5$, $n=1$ and $\sigma_8$=1.
This cosmology has an unrealistically high baryon fraction, a much smaller
acoustic oscillation scale ($\sim 50\,h^{-1}$Mpc compared with the
$\sim 100\,h^{-1}$Mpc of the `concordance' cosmology) and is more nonlinear
at $z=0$ than our Universe is believed to be.
This emphasizes the effects we are investigating while reducing the sampling
error, simplifying the numerical problem and allowing us to obtain highly
robust measures of small effects. In \S~\ref{sec:implications}, 
we extend the model constructed to concordance $\Lambda$CDM cosmologies, 
focusing on one with $\Omega_M=0.25$, $\Omega_B h^2=0.0224$, $h=0.72$, 
$n_s=0.97$ and $\sigma_8=0.8$ for definiteness.

\begin{figure}
\begin{center}
\resizebox{3in}{!}{\includegraphics{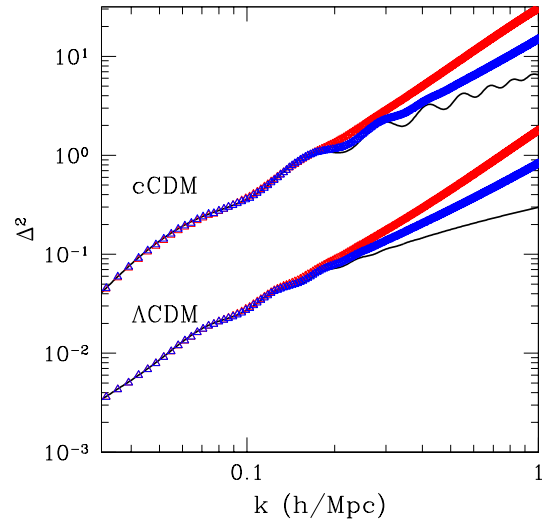}}
\end{center}
\vspace{-0.2in}
\caption{The nonlinear matter power spectrum at $z=0$ (red, upper) and $z=1$
(blue, middle), compared to the linear theory (black, lower) for our $c$CDM
and $\Lambda$CDM models.  The $z=1$ power spectra have been scaled by
$1/D^2$ to match the other power spectra on large scales and the $\Lambda$CDM
spectra have been offset (vertically) for clarity.  Note the strong damping
of the oscillations and the large excess power on small scales in the evolved
fields.}
\label{fig:pknonlin}
\end{figure}

To model nonlinear structure formation and the formation of dark matter
halos we used 10 independent simulations each of $1024^3$ particles in
periodic, cubical boxes of side length $2\,h^{-1}$Gpc.
The simulations were started at $z=100$ using the Zel'dovich approximation
and evolved to $z=0$ with the TreePM \cite{TreePM} code.
The full phase-space data were dumped at a number of redshifts between $z=1$
and $z=0$, and groups were found using the friends-of-friends algorithm
\cite{DEFW} with a linking length of 0.168 times the mean interparticle
spacing.  We keep all groups down to 10 particles, or
$2\times 10^{13}\,h^{-1}M_\odot$, using the sum of the particle masses in
the group as our halo mass definition for simplicity.  These minimum masses
correspond to peak heights running from $\nu\simeq 1$ at $z=0$ to $\nu\simeq 2$
at $z=1$.
The mass function and nonlinear power spectrum for this model are shown
in Figures \ref{fig:massfn} and \ref{fig:pknonlin} for reference.  Throughout
this paper we do not subtract shot-noise from the power spectra, but allow it to 
be a nuisance parameter in our fits (see below).
More details on these simulations, including convergence tests, are
in \cite{CarWhiPad09}.

Fitting the acoustic scale involves locating the position of the acoustic
feature while allowing for variations in the broad-band shape due to
e.g.~galaxy biasing. We do so by fitting the observed power spectra with
\begin{equation}
\label{eq:pfit}
P_{\rm fit}(k) = B(k) P_{\rm w}(k, \alpha) + A(k) \,\,,
\end{equation}
where $A(k)$ and $B(k)$ are smooth functions and $\alpha$ measures the 
acoustic scale relative to a ``best-guess'' cosmology. $P_{\rm w}$
is a template for the biased, nonlinear acoustic feature.
This definition of peak ``shift'' is over and above the shift in the point
where $\xi'(r)=0$, or shifts in the extrema of the oscillations in the power spectrum.
A good match between theory and observations, including the correct background
cosmology and hence distance-redshift relation, should give $\alpha\equiv 1$.

Note that the precise partitioning into acoustic feature and broad band shape 
is dependent on the particular choices of $A$ and $B$. Since constructing
an accurate template for the acoustic feature yields a good template 
for larger scales, we assume $B(k)$ is a constant. $A(k)$ is assumed
to be a cubic spline specified at $0.0$, $0.1$, $\cdots$, $0.4$ and 
derivatives specified at the end points. The shot noise component is 
simply absorbed into $A(k)$. The above prescription yields seven nuisance
parameters; we do not vary this (or the particular prescription) in this
paper, although we do get consistent results for different choices.

The fits are done by $\chi^2$ minimization, fitting the 70 power spectrum bins
between $k=0.02\,h\,{\rm Mpc}^{-1}$ and $0.35\,h\,{\rm Mpc}^{-1}$.
We assume a diagonal covariance matrix where the errors are a smooth fit to
the run to run variance of the 10 simulations.
The errors on all derived quantities are determined the variance of 1,000
bootstrap resamplings. 

We now turn to the purpose of this paper - the determination of $P_{\rm w}$,
first for the matter and then for biased tracers.

\begin{figure}
\begin{center}
\resizebox{3in}{!}{\includegraphics{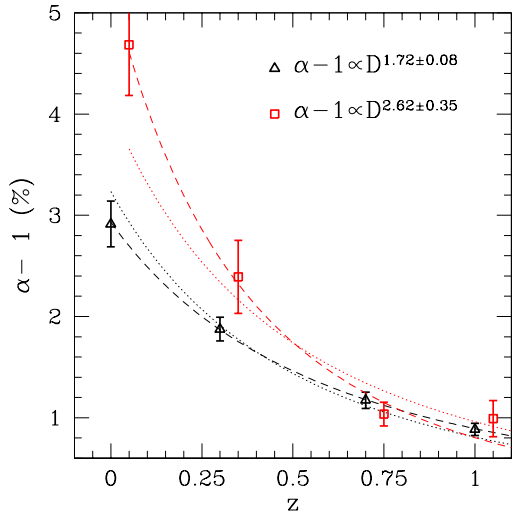}}
\end{center}
\vspace{-0.2in}
\caption{The shift in the acoustic scale, $\alpha-1$ vs.~redshift for the
mass (black triangles) and $\nu=1.9$ halos (red squares) using
Eq.~(\protect\ref{eqn:ESW}) as the template. 
Also shown are the best fit power-laws (dashed) and the expectations of
perturbation theory [$\alpha-1\propto D^2(z)$] (dotted).
The points and curves for the halos are shifted by $\delta z = 0.05$ for clarity.
}
\label{fig:alpha_vs_z}
\end{figure}

\section{Matter}

As is dramatically evident in Fig.~\ref{fig:pknonlin},
the sharp acoustic feature at high redshift gets smeared
by bulk flows and super-cluster formation as the Universe evolves
\cite{Bha96a,Bha96b,MeiWhiPea99,ESW07,CroSco08,Mat08}.
An estimate of the smearing is given by convolving the matter correlation
function with a Gaussian of width equal to the rms displacement of particles
from their initial positions.  To lowest order this is the rms Zel'dovich
displacement
\begin{equation}
  \Sigma_1^2 = \frac{1}{3\pi^2} \int dk P_L(k)
\label{eqn:Sigma}
\end{equation}
suggesting a template of the form
\begin{equation}
  P_{\rm w}(k, \alpha) = \exp\left(-\frac{k^2\Sigma^2}{2}\right) P_L(k/\alpha)
\label{eqn:ESW}
\end{equation}
where $P_L$ is the linear theory power spectrum at the redshift of interest,
and $\Sigma$ is allowed to vary.
The lowest order result is the same whether obtained using the
peak-background split \cite{ESW07},
renormalized perturbation theory \cite{CroSco08} or
resummed Lagrangian perturbation theory \cite{Mat08}.
While the Gaussian form of the smearing is a reasonable approximation to what
we see in the simulations, we find $\mathcal{O}(10\%)$ deviations from
$\Sigma_1$ in our cosmology.

Figure~\ref{fig:alpha_vs_z} and Table~\ref{tab:dmfits} summarize the shifts
obtained using the above template. We find that $\alpha \ne 1$ at very high
significance, with $\alpha-1\sim 4\%$ at $z=0$.
We remind the reader that this is an extreme cosmology; the shifts for a
concordance cosmology (as we discuss in \S\ref{sec:implications}) are
approximately an order of magnitude smaller. 
We also observe that the shifts decrease with redshift as $\alpha-1 \sim D^2$
(adopting the convention $D(z=0) = 1$),
consistent with the field getting more linear at higher redshift and suggestive
that the next order terms in perturbation theory
(${\cal O}(P_{L}^2)$, see  \cite{CarWhiPad09} for a recent review)
are responsible (see also \cite{CroSco08,SSS08} who emphasized this point).

\begin{table}
\begin{tabular}{c|ccc}
$z$ & DM & x$\delta_L$ & w/ $P_{22}$ \\
\hline
0.0  & $\phantom{-} 2.91 \pm 0.20$ & $- 0.19 \pm 0.08$ & $- 0.03 \pm 0.16$\\
0.3  & $\phantom{-} 1.88 \pm 0.12$ & $- 0.18 \pm 0.11$ & $- 0.38 \pm 0.09$\\
0.7  & $\phantom{-} 1.17 \pm 0.07$ & $- 0.13 \pm 0.11$ & $- 0.12 \pm 0.05$\\
1.0  & $\phantom{-} 0.88 \pm 0.06$ & $- 0.11 \pm 0.12$ & $- 0.04 \pm 0.04$\\
\hline
\end{tabular}
\caption{\label{tab:dmfits} Shifts, i.e.~$\alpha-1$ in per cent, for the
matter density autopower spectrum and the cross-spectrum of the density
with the linearly extrapolated initial density field. The presence of a
shift in the first column and not in the second demonstrates that the shifts
arise from the higher order $P_{mn}$ ($m,n \ge 2$) contributions to the power spectrum
(see text for details).
The last column demonstrates that this shift can be corrected by adding a
$P_{22}$ term to the template.}
\end{table}

To explore this possibility, we expand the
density contrast in powers of linear density $\delta_{L}$
\begin{equation}
  \delta = \delta^{(1)} + \delta^{(2)} + \delta^{(3)} + \cdots
\end{equation}
with $\delta^{(1)}\equiv\delta_L$.  It is straightforward to show that
\begin{eqnarray}
  \delta^{(n)}(k) &=& \int \frac{d^3q_1\ldots d^3q_n}{(2\pi)^{3n}}
  \ (2\pi)^3\delta_D\left(\sum q_i-k\right) \nonumber \\
  &\times& F_n(\{q_i\},k)\ \delta_L(q_1)\ldots\delta_L(q_n)
\end{eqnarray}
where the $F_n$ contain dot products of the vectors $q_i$ and can be
generated from recurrence relations \cite{Goroff86,Makino92,Jain94}.
If the initial field is Gaussian the nonlinear power spectrum is then
given by
\begin{equation}
P_{NL} = \left\{P_{11} + P_{13} + P_{15} + \cdots\right\}
       + \left\{P_{22} + \cdots \right\}
\end{equation}
where $P_{ij} = \langle \delta^{(i)} \delta^{(j)} \rangle$ and $P_{11}=P_L$.
In what follows, we refer to these groups of terms as the $P_{1n}$ and
$P_{mn}$ terms respectively.
The $P_{1n}$ terms also arise if we take the cross-spectrum of the initial
and evolved fields, while the $P_{mn}$ terms only arise in
the auto-spectrum.

\begin{figure}
\begin{center}
\resizebox{3in}{!}{\includegraphics{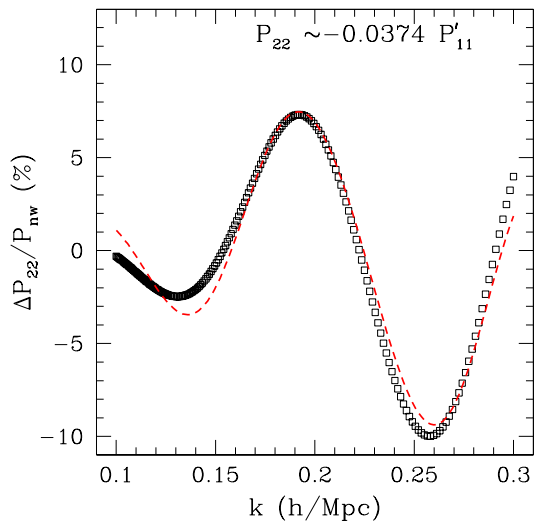}}
\end{center}
\vspace{-0.2in}
\caption{The out-of-phase contribution predicted by perturbation theory well
approximates the derivative of the acoustic signal.  The points plot $P_{22}$,
while the line is a scaled version of $dP_L/d\ln k$ with the scaling given in
the inset. The smooth components of both curves have been subtracted by fitting
a cubic spline to the data. All curves are for $z=0$.}
\label{fig:derivP00_M004}
\end{figure}

Ignoring combinatorial factors, the $P_{1n}$ terms are given by
\begin{equation}
P_{1n}(k) \sim P_{L}(k) \int \prod_{k=1}^{(n-1)/2} \left[ d^{3} q_{k} P_{L}(q_{k}) \right]
F_{n}(\cdots) \,\,.
\end{equation}
If we factor out the common $P_L$ and focus on the lowest order correction,
$P_{13}$, we see that it involves a integral over a single $P_L$ and a
relatively broad kernel which suppresses the oscillations.
We expect these terms not to lead to significant out-of-phase contributions,
though they can contribute corrections to the damping described above.
Since we can isolate these terms by considering the cross-spectrum between
the initial and evolved matter fields, we can test the above hypothesis.
Table~\ref{tab:dmfits} demonstrates that the shifts in this cross-spectrum
are reduced by over an order of magnitude compared to the auto-spectrum
and are consistent with zero given our statistical precision.
Note that the above argument is only true for the lowest order
contribution, but higher order terms are suppressed by additional factors
of ${\cal O}(\delta^2)$, and therefore drop off even more strongly with
redshift.

\begin{figure*}
\begin{center}
\resizebox{2in}{!}{\includegraphics{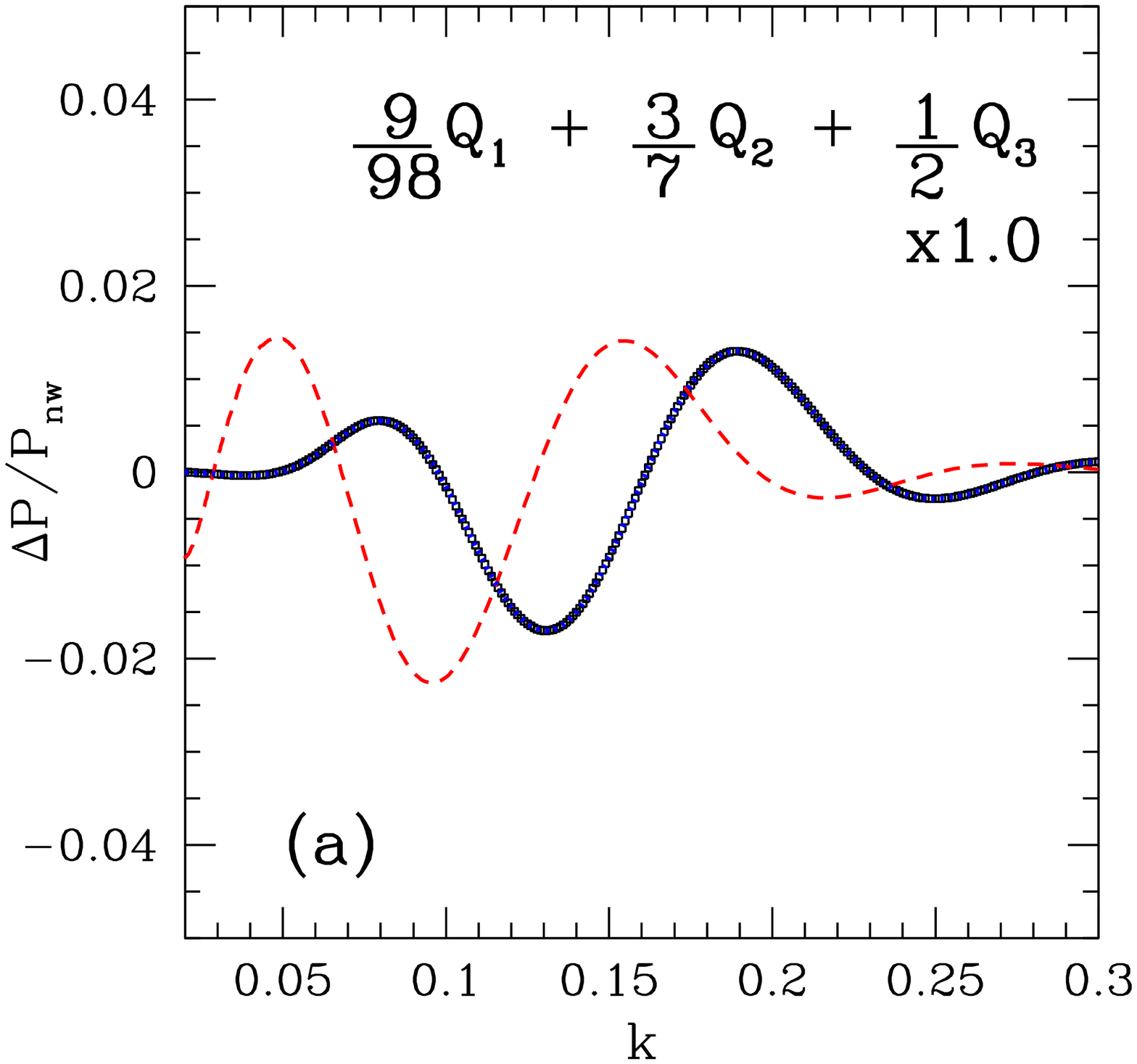}}
\resizebox{2in}{!}{\includegraphics{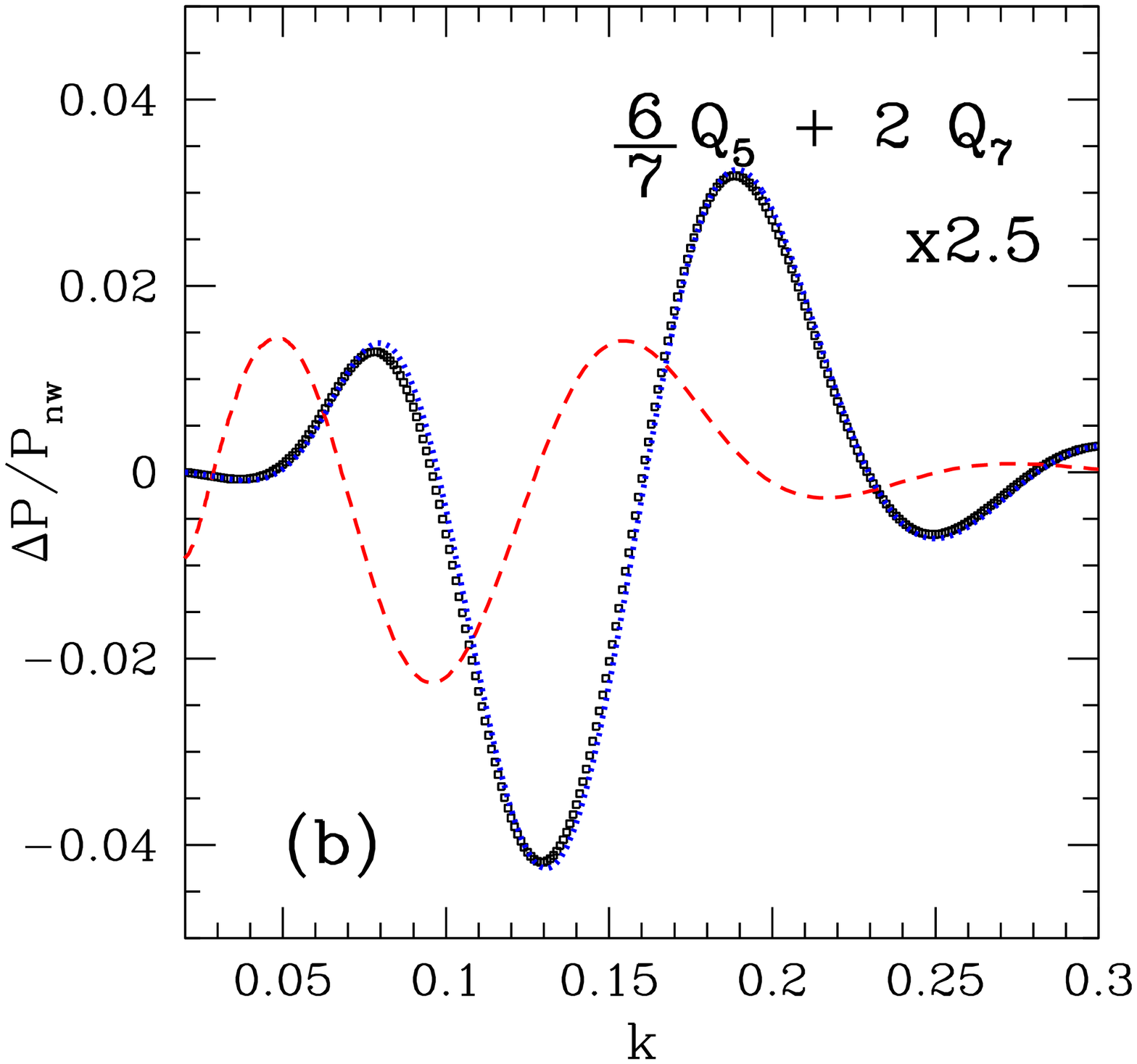}}
\resizebox{2in}{!}{\includegraphics{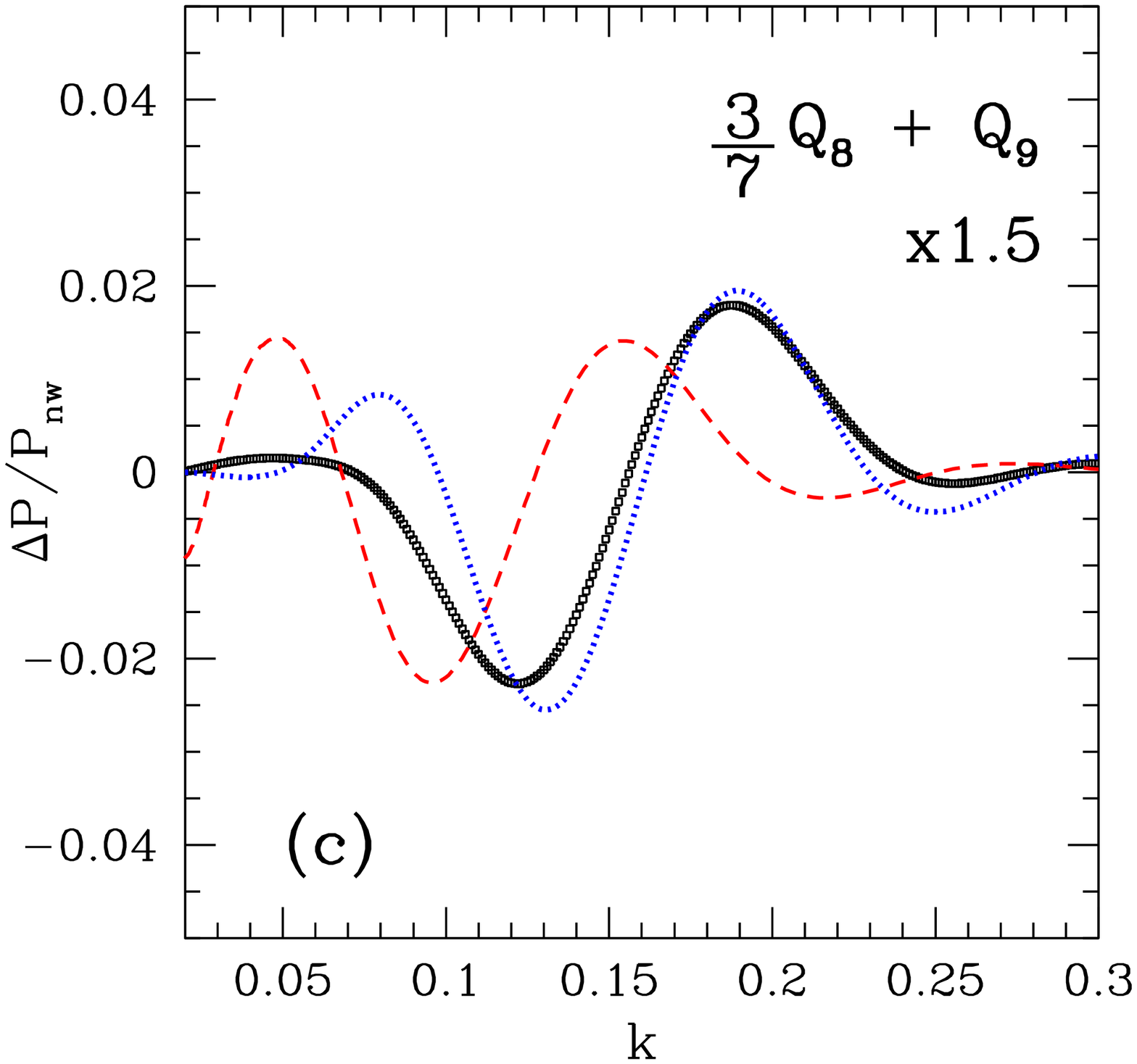}}
\resizebox{2in}{!}{\includegraphics{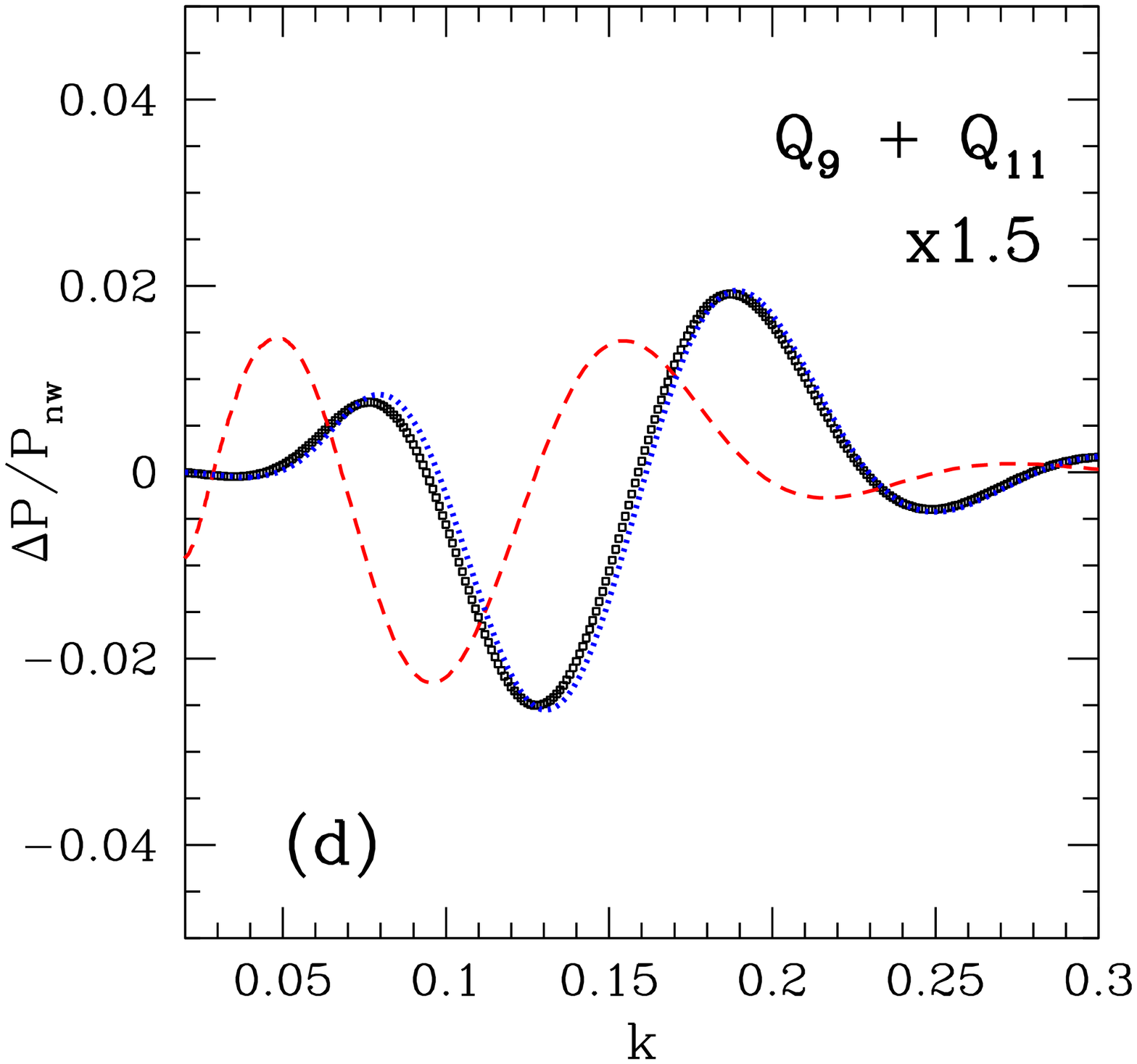}}
\resizebox{2in}{!}{\includegraphics{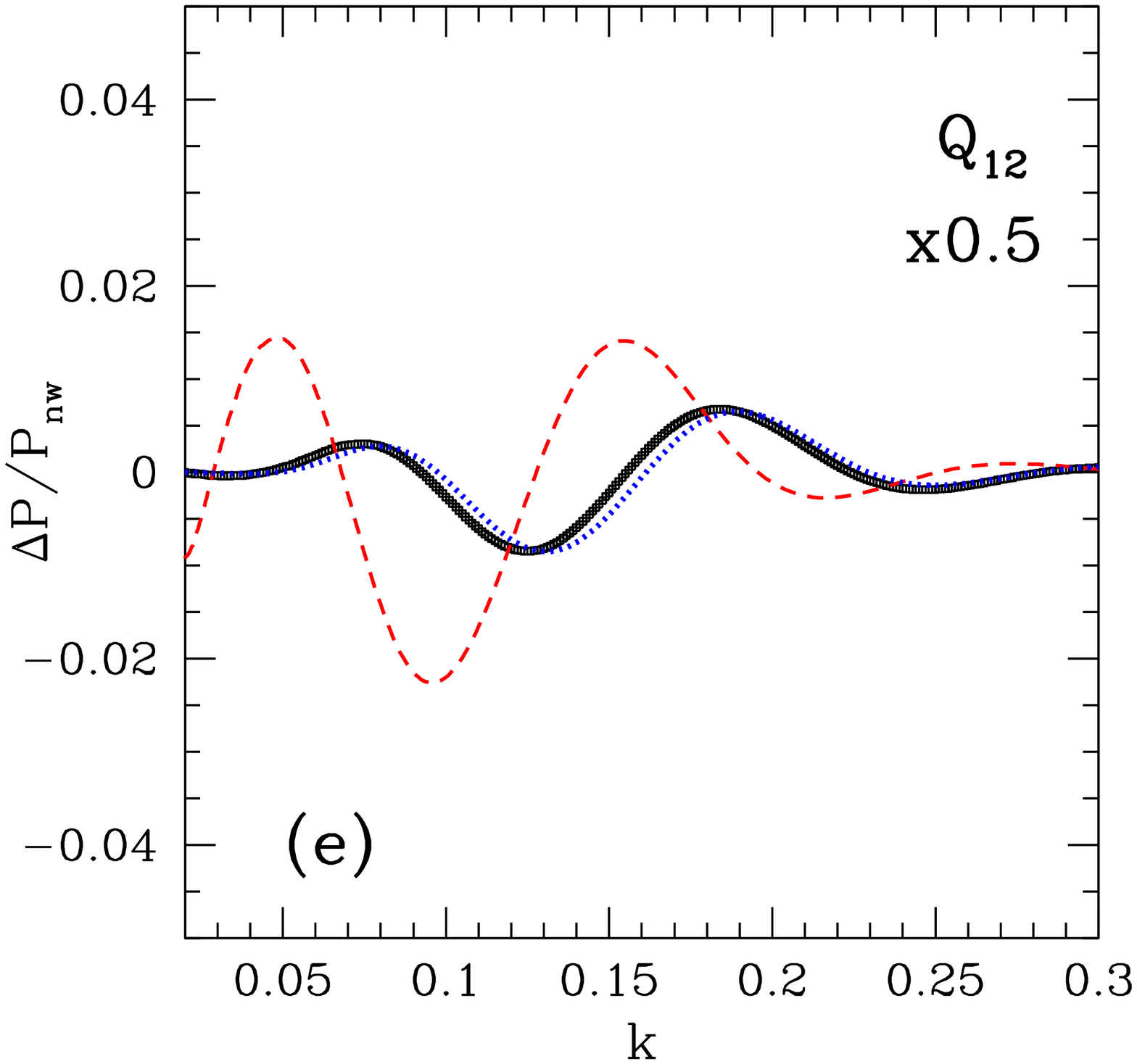}}
\resizebox{2in}{!}{\includegraphics{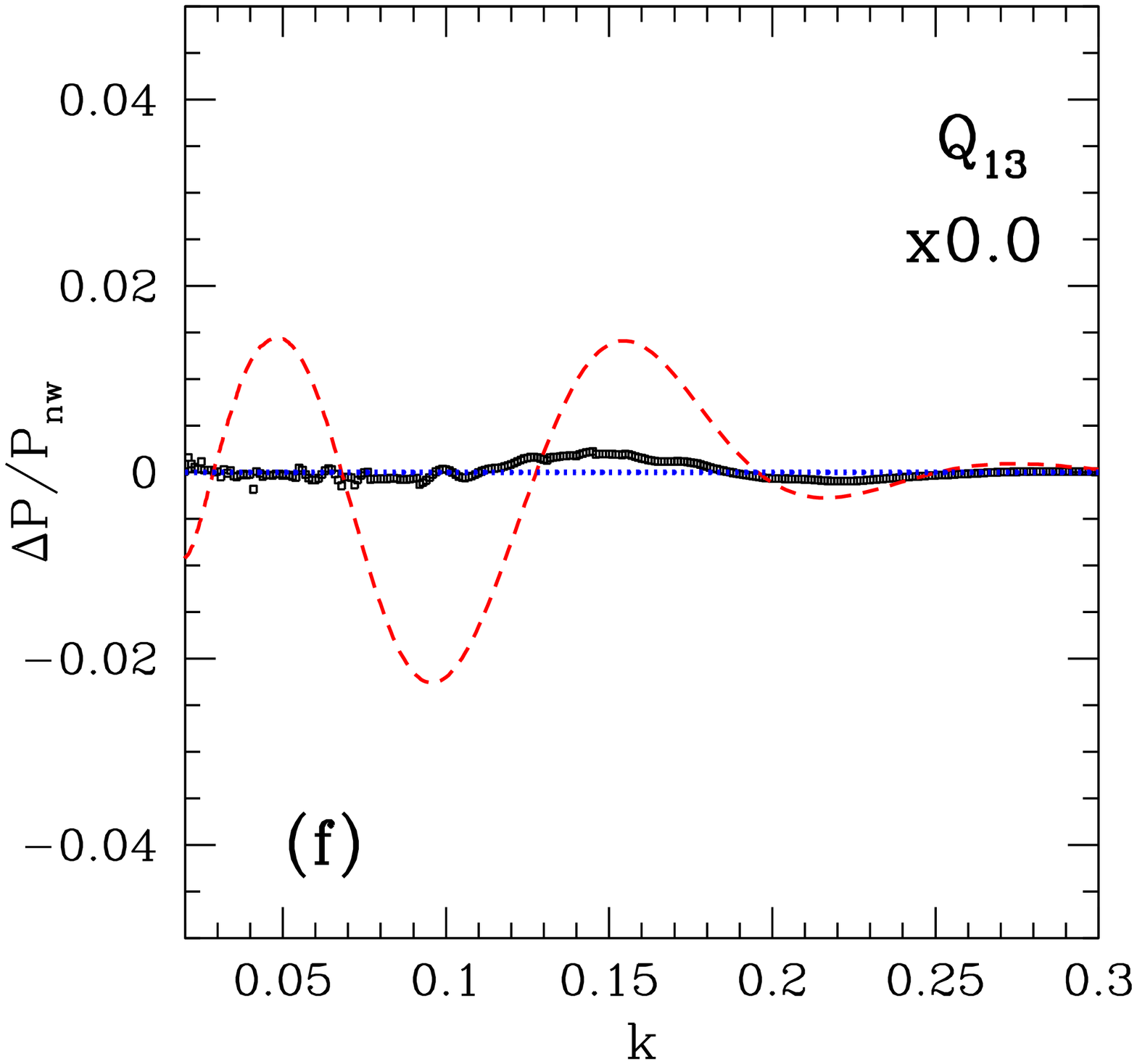}}
\end{center}
\vspace{-0.2in}
\caption{The six combinations of $Q_{n}(k)$, including the exponential 
damping (at $z=0$) that appear in the nonlinear halo power spectrum for $c$CDM.  
For each of these, we subtract a smooth component
(by fitting a five point cubic spline), and compare to the ``no-wiggle''
power spectrum of \protect\cite{EisHu98}.
The dashed [red] line shows the same procedure applied to the linear power
spectrum (divided by 5), while the dotted [blue] line is the $P_{22}$ correction to the
nonlinear matter power (upper left panel).  In each panel we have scaled
$P_{22}$ by the multiplicative factor shown to better match each combination
of $Q_n$.}
\label{fig:qs4}
\end{figure*}

The lowest order $P_{mn}$ term is 
\begin{equation}
P_{22}(k) = \frac{9}{98}Q_{1}(k) + \frac{3}{7}Q_{2}(k)
              + \frac{1}{2} Q_{3}(k) \,\,,
\end{equation}
where the $Q$s are defined in Appendix~\ref{sec:qdefs}. In constrast
with $P_{13}$, these terms (see Eq.~\ref{eq:qndef}) involve integrals
of products of $P_L$ which can lead to out-of-phase terms.
For example, $F_2$ is peaked around $q_1\approx q_2\approx k/2$.  When
$P_L$ contains an oscillatory piece, e.g.~$\sin(kr)$, $P_{22}$ contains
a piece schematically of the form $\sin^2(kr/2)\sim 1+\cos(kr)$, which
oscillates out-of-phase with $P_L$.
Panel (a) of Fig.~\ref{fig:qs4} explicitly shows this; in fact, this
oscillatory part of $P_{22}$ is very similar to scaled log-derivative
of $P_{L}$ (Fig.~\ref{fig:derivP00_M004}; see also \cite{CroSco08}).
Taylor expanding a shifted power spectrum, 
\begin{equation}
\label{eq:taylorshift}
  P_L(k/\alpha)\simeq P_L(k) - \left(\alpha-1\right)\frac{dP_L}{d\ln k}
  + \cdots \,
\end{equation}
we find good agreement between the predicted shift of Fig.~\ref{fig:derivP00_M004}
and the measured shift in Table~\ref{tab:dmfits}.
It is important to note that we have subtracted smooth components for all
of these comparisons suggesting that even though perturbation theory
does not accurately predict the broad-band shape \cite{CarWhiPad09}, it 
does capture the evolution of the acoustic feature.

The above suggests a modified template, 
\begin{eqnarray}
\label{eq:pw2}
P_{\rm w}(k, \alpha) &=&  \exp\left(-\frac{k^2\Sigma^2}{2}\right) P_L(k/\alpha) \nonumber \\
&+& \exp\left(-\frac{k^2\Sigma_{1}^2}{2}\right) P_{22}(k/\alpha) \,.
\end{eqnarray}
Note that we have damped the oscillations in $P_{22}$ as for $P_{L}$ although
we fix the damping scale to the first order calculation. While this damping
follows naturally from the heuristic picture described at the start of this section, 
it is also a consequence of the resummations in eg. Lagrangian perturbation theory.
Table~\ref{tab:dmfits} shows that such a template corrects for the shifts observed in the 
matter. 

\begin{figure}
\begin{center}
\resizebox{3in}{!}{\includegraphics{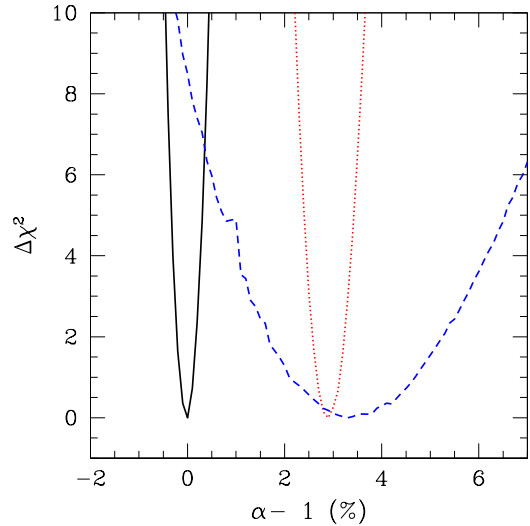}}
\end{center}
\vspace{-0.2in}
\caption{The $\Delta\chi^2$ for fits of our $z=0$ mass power spectrum to the
functional form of Eq.~\protect\ref{eqn:ESW} (dashed red), to the form
including $P_{22}$ (solid black) and marginalizing over the amplitude of the
$P_{22}$ term (dashed blue).  There are $60$ degrees of freedom in the fit,
and in each case the best fit is a reasonable fit.
Note that Eq.~\protect\ref{eqn:ESW} gives a biased acoustic scale,
including the $P_{22}$ term eliminates the bias, and allowing the amplitude
of the $P_{22}$ term to float results in very weak constraints.}
\label{fig:chi2}
\end{figure}

Looking ahead to biased tracers, we note that one cannot simultaneously fit
for the amplitude of the $P_{22}$ term and the shift, since these are 
highly degenerate (Eq.~\ref{eq:taylorshift}). Doing so results in highly
degraded constraints on the acoustic scale, as is evident in Fig.~\ref{fig:chi2}. 
We need to know the relative amplitude of $P_{22}$ and $P_{11}$ to correct
the shift.  For the mass the relative amplitude is straightforwardly given
by perturbation theory.  Do the same terms come in for biased tracers and
are we able to determine the relative amplitude of the two types of terms?

\section{Halos}

\subsection{Shifts}

We investigate the acoustic signal of biased tracers in our simulations by
computing the clustering of samples of dark matter halos chosen to lie in
narrow mass ranges.
Specifically we use the linear theory power spectrum to convert from halo
mass to peak height, $\nu\equiv\delta_c/\sigma(M)$, and pick halos in the
range $0.85\le\nu<1.15$, $1.15\le\nu<1.45$, $\cdots$.  For these we compute
both the auto-power spectrum and the
cross-power spectra with the linear and evolved dark matter density field.
Given the low number density of most of our samples, we focus on the 
cross-power spectra in the analysis below.

\begin{table}
\begin{tabular}{c|cccc}
$z$ & $\nu=$1.0 & 1.3 & 1.6 & 1.9 \\
\hline
0.0  & $\phantom{-} 1.66 \pm 0.19$ & $\phantom{-} 1.91 \pm 0.33$ & $\phantom{-} 2.20 \pm 0.36$ & $\phantom{-} 4.68 \pm 0.51$\\
0.3  & -  & $\phantom{-} 1.38 \pm 0.17$ & $\phantom{-} 1.53 \pm 0.26$ & $\phantom{-} 2.39 \pm 0.36$\\
0.7  & -  & -  & $\phantom{-} 0.85 \pm 0.09$ & $\phantom{-} 1.04 \pm 0.15$\\
1.0  & -  & -  & -  & $\phantom{-} 0.99 \pm 0.17$\\
\hline
0.0  & $- 0.10 \pm 0.06$ & $- 0.23 \pm 0.07$ & $- 0.16 \pm 0.10$ & $- 0.10 \pm 0.12$\\
0.3  & -  & $- 0.01 \pm 0.06$ & $- 0.20 \pm 0.08$ & $\phantom{-} 0.01 \pm 0.06$\\
0.7  & -  & -  & $- 0.03 \pm 0.06$ & $- 0.10 \pm 0.04$\\
1.0  & -  & -  & -  & $- 0.04 \pm 0.05$\\
\hline
0.0  & $- 0.01 \pm 0.15$ & $\phantom{-} 0.12 \pm 0.25$ & $- 0.19 \pm 0.26$ & $\phantom{-} 0.51 \pm 0.31$\\
0.3  & -  & $- 0.20 \pm 0.11$ & $- 0.37 \pm 0.17$ & $- 0.58 \pm 0.23$\\
0.7  & -  & -  & $- 0.27 \pm 0.07$ & $- 0.36 \pm 0.10$\\
1.0  & -  & -  & -  & $- 0.13 \pm 0.11$\\
\hline
\end{tabular}
\caption{\label{tab:shiftsDM} Shifts in the measured acoustic scale probed
by the halo density field as a function of halo peak height and redshift.
The first group of numbers are the measured shifts for the halo density field
correlated with the matter field, fit with the template of Eq.~\ref{eqn:ESW}.
The second group are for the halo density correlated with the linear density
field, again fit with the template of Eq.~\ref{eqn:ESW}.
The final set of numbers are analogous to the first, except fit using the
template including the $P_{mn}$ corrections.
Note that the shifts present in the first group are significantly reduced
in the other groups.}
\end{table}

Motivated by the development in the previous section, we test if analogous
results exist for halos; Table~\ref{tab:shiftsDM} and Fig.~\ref{fig:alpha_vs_z}
summarize our findings. We find that (i) the halos exhibit non-zero shifts
that are functions of halo type, (ii) the shifts scale  approximately as $D^2$,
and (iii) the shifts are once again absent in the cross-spectrum with the linear
density field. As with the matter, this argues
that, within the language of perturbation theory, the
shifts come from $P_{mn}$ terms and are dominated by second order corrections.
The amplitude of the $P_{mn}$ terms relative to the $P_{11}$ terms depends
on the type of tracer, to which we now turn.

\subsection{Eulerian and Lagrangian Bias}

We can proceed to develop the perturbation theory of biased tracers in two
ways: via Eulerian or Lagrangian perturbation theory.  We begin with the
former and follow \cite{FryGaz93} in defining
\begin{equation}
\label{eq:eulerb}
  \delta_h = b_1^E \delta + \frac{b_2^E}{2!}\delta^2 + \cdots \,.
\end{equation}
Implicit here is that the halo density is defined in configuration space,
and that the density fields have been smoothed on some scale $R$ to
allow us to truncate the expansion. We assume that we are working on scales
$\gg R$, and will ignore subtleties that arise from the smoothing 
\cite{HeaMatVer98,PatBias06}
for now;  we explicitly reinstate the smoothing scale in Sec.~\ref{sec:empirical_b1b2}.

The halo auto-power spectrum in Eulerian perturbation theory becomes
(see also \cite{HeaMatVer98,PatBias06,SSS07})
\begin{eqnarray}
  P_h &=& \left(b_1^E\right)^2\left( P_{11} + P_{22} \right)
      + b_1^Eb_2^E\left( \frac{3}{7}Q_8+Q_9 \right) \nonumber \\
     &+& \frac{(b_2^E)^2}{2} Q_{13} + \cdots
\end{eqnarray}
with terms like $P_{1n}$ included in the missing terms denoted $\cdots$.
The cross spectrum between two tracers $I$ and $II$ can be obtained by
the replacements
\begin{eqnarray}
\label{eq:crossmap}
  b_n   &\to& \frac{1}{2}\left( b_n^{(I)}+b_n^{(II)} \right) \nonumber \\
  b_n^2 &\to& b_n^{(I)}b_n^{(II)} \nonumber \\
  b_1b_2&\to& \frac{1}{2}\left( b_1^{(I)}b_2^{(II)}+b_1^{(II)}b_2^{(I)}\right)
\end{eqnarray}
As an example, for the cross-spectrum with the mass we obtain
\begin{equation}
  P_{h,m} = b_1^E\left( P_{11} + P_{22} \right)
     + \frac{b_1^Eb_2^E}{2} \left( \frac{3}{7}Q_8+Q_9 \right) + \cdots
\end{equation}
Figure \ref{fig:qs4} shows that the $P_{22}$ and $\frac{3}{7}Q_8+Q_9$ terms
contain out-of-phase oscillations with very similar shapes while the $Q_{13}$
term is essentially non-oscillatory.  This suggests that biased tracers will
exhibit different shifts than the matter, and the difference will
depend on the structure of the bias.  It is also worth pointing out that the
shift for a $b_1\equiv 1$ tracer is {\it not\/} the same as for the mass -
a fact also evident in Table~\ref{tab:shiftsDM} where the $\nu=1$ halos exhibit
different shifts from the matter!

The alternative description is within the Lagrangian picture, which has
recently been developed in \cite{MatsubGal} (see also Appendix~\ref{sec:d2d_LPT} for the basic
definitions).  Within this formalism the
halo auto-spectrum can be written
\begin{eqnarray}
  P_h &=& \exp\left[-\frac{k^2\Sigma^2}{2}\right]
  \left\{ \left(1+b_1^L\right)^2 P_{11} + P_{22}
  \vphantom{\int} \right. \nonumber \\
  &+& b_1^L\left[\frac{6}{7}Q_5+2Q_7\right]
   + b_2^L\left[\frac{3}{7}Q_8+Q_9\right] \nonumber \\
  &+& \left(b_1^L\right)^2\left[Q_9+Q_{11}\right] \nonumber \\
  &+& 2b_1^L b_2^L Q_{12}
   +  \left. \frac{1}{2}\left(b_2^L\right)^2 Q_{13} \right\} + \cdots
\end{eqnarray}
where again terms like $P_{1n}$ have been included in $\cdots$. 
As before, expressions for cross-spectra follow from the mapping 
in Eq.~\ref{eq:crossmap}, taking care to expand the $(1+b^L_1)^2$ term
before making the substitutions.
Again, Figure \ref{fig:qs4} shows that the terms which arise look like
scaled versions of $P_{22}$, except for $Q_{13}$ which is non-oscillatory.

\begin{figure}
\begin{center}
\resizebox{3in}{!}{\includegraphics{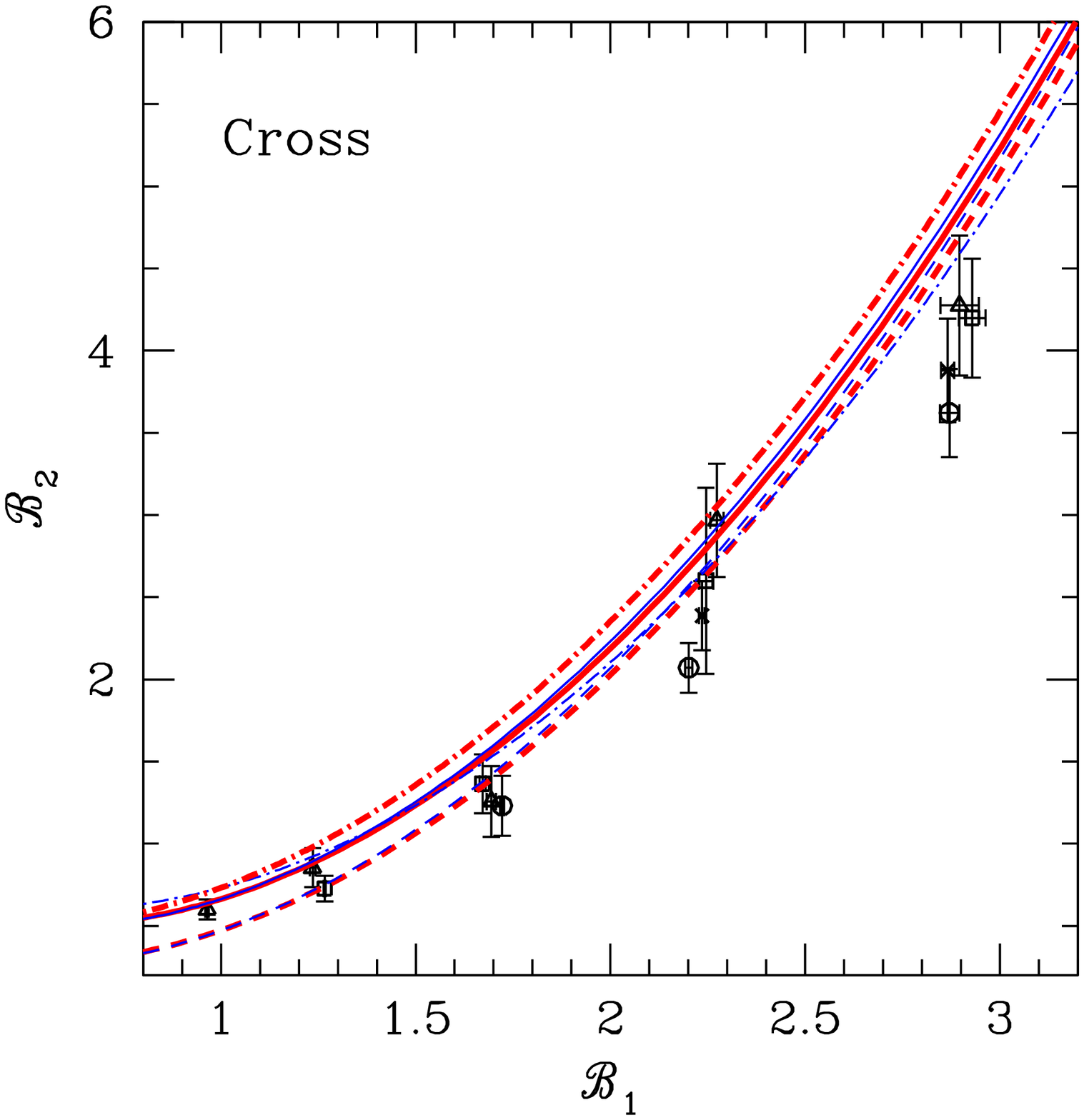}}
\resizebox{3in}{!}{\includegraphics{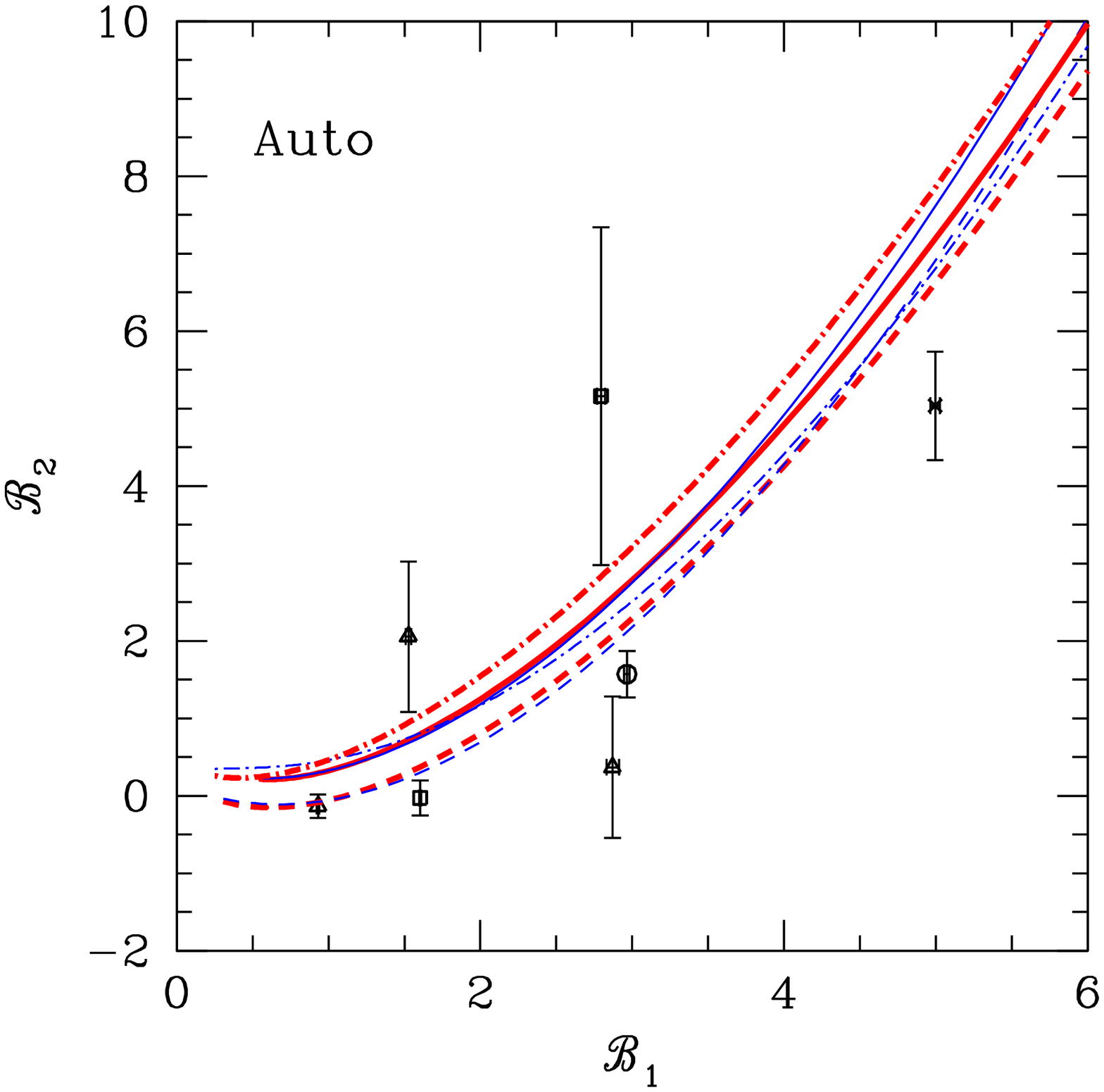}}
\end{center}
\vspace{-0.2in}
\caption{(Upper) The best fit values of ${\cal B}_{1}$ and ${\cal B}_{2}$ for
the cross power spectra of halos and the evolved matter density; triangles,
squares, circles, and crosses are the $z=0$, $0.3$, $0.7$ and $1$ data
respectively.  Note that increasing ${\cal B}_{1}$ corresponds to increasing $\nu$.
The solid lines are Sheth-Tormen \protect\cite{STbias}
predictions, while the dashed lines are for Press-Schecter \protect\cite{PS};
the thick [red] lines are for Lagrangian theory, while the thin [blue]
lines are for Eulerian theory. The dot-dashed lines are based on a quadratic fit
to $b_2(b_1)$, again both for Eulerian and Lagrangian bias models.
The theoretical scatter in these relations is $\sim 20\%$.
(Lower) As above but for the halo-halo auto-spectrum.
}
\label{fig:b1b2}
\end{figure}

The structure of these two sets of power spectra appears quite different,
especially in the scaling of the different out-of-phase terms with $b_n$.
However once broad-band power is removed, both of these cases can be
effectively written as 
\begin{equation}
\label{eq:Phalo_eff}
  P_h =  \exp\left(-\frac{k^2 \Sigma^{2}}{2}\right)
    \left[{\cal B}_{1}P_{L}  + {\cal B}_{2}P_{22}\right] .
\end{equation}
where we have implicitly assumed that in both cases some of the higher
order terms we have neglected above would sum to an exponential damping,
as happens in some variants of both Eulerian and Lagrangian perturbation
theory.
For the autopower spectra, using the empirically determed scalings in
Figure \ref{fig:qs4}, the ${\cal B}$s are related to the bias
parameters by
\begin{equation}
{\cal B}_1  =  (b^{E}_{1})^2\ ,  \qquad
{\cal B}_2 = (b^{E}_{1})^2 + \frac{3}{2} b^{E}_{1} b^{E}_2 \,
\end{equation}
or
\begin{eqnarray}
\label{eq:b2calb}
{\cal B}_1 & = & (1 + b^{L}_{1})^2 \, \nonumber \\
{\cal B}_2 & = & 1 + \frac{5}{2} b^L_1 + \frac{3}{2} b^L_2 +
                 \frac{3}{2}(b^L_1)^2 + b^L_1 b^L_2 \,. 
\end{eqnarray}
Analogous expressions can be written for the cross-spectra by making the
substitutions described above.

If we fit our N-body data to Eq.~(\ref{eq:Phalo_eff}) we find that this
form is a good description of the data and the different samples all lie
in a narrow band in the ${\cal B}_1-{\cal B}_2$ plane, as shown in
Figure \ref{fig:b1b2} for both the auto- and cross-spectra.
This suggests that the halos form a 1-parameter family in terms of the
nonlinear bias.
This is fortunate, because the ratio of ${\cal B}_2$ to ${\cal B}_1$
is degenerate with the shift of the acoustic scale.

\subsection{Explaining the Shifts}

The previous results simply imply that the amplitude of the shifts is 
a function of halo bias.
However, the perturbative formulation of the previous section also relates
the amplitude of the shift to the bias parameters of the halos.
We test this relationship here, first using the peak-background split model
(see e.g.~\cite{ColKai89,MoWhite,STbias}) and an empirically calibrated
$b_1$-$b_2$ relationship from simulations.

\subsubsection{Peak-Background Split}

The starting point for the peak-background split is the unconditional
multiplicity function
\begin{equation}
 \nu f(\nu)\,d\nu = \frac{M}{2\bar{\rho}}\ \frac{dn}{dM}\, dM
\end{equation}
which can be fit with
\begin{equation}
\label{eq:stmass}
  \nu f(\nu) \propto \left(1 + \frac{1}{(a\nu^{2})^{p}}\right)
  \left(\frac{a\nu^2}{2}\right)^{1/2}
  \exp\left(- \frac{a\nu^{2}}{2}\right)
\end{equation}
where $a=1$, $p=0$ gives the Press-Schecter mass function \cite{PS},
while $a=0.707$, $p=0.3$ yields the Sheth-Tormen mass function \cite{STbias}.
Within the assumption of the peak-background split, the conditional
multiplicity function is given by the substitution, 
\begin{equation}
  \nu \rightarrow \nu \left(1 - \frac{\delta}{\delta_{c}}\right) \,\,,
\end{equation}
where $\delta$ is the background density and $\delta_c\simeq 1.686$ is the
critical overdensity for collapse.
The Lagrangian bias parameters then follow from Taylor expanding the
(appropriately normalized) conditional multiplicity function as a function
of $\delta$, yielding $b^L_n=[\nu f(\nu)]^{-1} d^n/d\delta^n[\nu f(\nu)]$ or
\begin{equation}
\label{eq:blag1}
  b^{L}_{1}(\nu) = \frac{1}{\delta_c} \left[ \nu^{2} - 1
  + \frac{2p}{1 + (a\nu^{2})^{p}}\right] \,, 
\end{equation}
and 
\begin{equation}
\label{eq:blag2}
  b^{L}_{2}(\nu) = \frac{1}{\delta_c^2} \left[a^2 \nu^4 -3 a \nu^2 + 
       \frac{2p(2a\nu^2 + 2p - 1)}{1 + (a\nu^{2})^{p}} \right]\,\,.
\end{equation} 
The Eulerian bias parameters are then defined by the mapping of the halo 
density $\delta_h$ from Lagrangian to Eulerian space,
\begin{equation}
  \delta_h^E = (1+ \delta^{E}) (1 + \delta_{h}^{L}) - 1
\end{equation}
where the factor of $(1+\delta^{E})$ comes from the mapping of the Lagrangian
to Eulerian volumes.  Assuming that the Eulerian and Lagrangian densities may
be related by the Taylor series \cite{MoWhite}
$\delta^L = \delta^E + c (\delta_{E})^2$
with $c\simeq -0.805$ assuming spherical collapse, we obtain
(see also \cite{Sco01})
\begin{equation}
\label{eq:beul1}
  b^{E}_1 = 1 + b^{L}_1 \,,
\end{equation}
and 
\begin{equation}
\label{eq:beul2}
  b^E_2 = b^L_2 + 2b^L_1 (1 + c) \,\,.
\end{equation}
 
The theory then makes predictions for both the mass-halo cross-spectrum and
the halo-halo auto-spectrum, with the former having significantly less shot
noise.
Fig.~\ref{fig:b1b2} compares these predictions to the observed relation
between the ${\cal B}$s, where we have translated from the bias parameters
using the prescription in the previous section.
For the cross-spectrum the simple model describes the observed trend with ${\cal B}_1$,
with the differences between the Press-Schecter and Sheth-Tormen predictions
being small (and not distinguishable by the data). Both models overpredict
${\cal B}_2$ at high ${\cal B}_1$.
Interestingly, the predictions for the Eulerian and Lagrangian descriptions
are virtually indistinguishable, even though the structure of the expressions
for the power spectra are very different.
For the auto-spectrum we have much more limited N-body data, but again the
simple model does an adequate, if not perfect, job of describing what we see.

\subsubsection{An empirical $b_1 - b_{2}$ relation}
\label{sec:empirical_b1b2}

A second approach is to empirically calibrate $b_1$ and $b_2$ using models
based upon simulations.  Recall that in any survey aiming to measure BAO
there will be ample data on small scales with which to construct models of
the tracers.

The traditional approach to determining the Eulerian bias parameters is to
compare various moments of ``counts-in-cells'' to the perturbative expressions.
A concern with such an approach is the effective scale of the measurements,
and the validity of the perturbative expressions.
An alternative is to compare the auto-spectra of the halos and the mass with
the cross-spectrum.  The combination of the three spectra can be used to
isolate $b_1$ and $b_2$, but this tends to be very noisy and proper shot-noise
subtraction is an issue.
We outline a different approach below that explicitly only uses large scales;
we defer detailed comparisons with other methods to future work.

We start by considering the configuration space statistic 
\begin{eqnarray}
{\cal S}(\bx) & = & \langle \delta_{L}^{2}(\bx_1) \delta_{h} (\bx_1)\rangle \nonumber \\
& = & b^E_1\langle \delta_L^2(\bx_1) \delta_S(\bx_2) \rangle
+ \frac{b^E_2}{2} \langle \delta_L^2(\bx_1) \delta_S^2(\bx_2) \rangle \,\,,
\end{eqnarray}
where $\bx = \bx_1-\bx_2$ and 
$\delta_S$ is the nonlinear matter density smoothed on a scale $R$ 
such that the Eq.~\ref{eq:eulerb}
is valid. Working to second order in the density field, the second term above 
reduces to $b_2 \langle \delta_L(\bx_1) \delta_S(\bx_2) \rangle^2$. If we work on
large scales $|\bx_1 - \bx_2| \gg R$, then we can approximate the smooth fields
by the underlying density field. Fourier transforming, we obtain 
\begin{equation}
\label{eq:Sk}
  {\cal S}(k) = \frac{b^{E}_{1}}{2} \left(\frac{3}{7} Q_{8} + Q_{9}\right)
  + b^{E}_{2} Q_{13} \,\,. 
\end{equation}
The $k \rightarrow 0$ limit yields a direct measure of $b^{E}_{2}$,
\begin{equation}
{\cal S}(k \rightarrow 0) = b^{E}_{2} Q_{13}(0)
\end{equation}
where 
\begin{equation}
Q_{13}(0) = \int \frac{d^{3}q}{(2\pi)^{3}} \left[P_{L}(q)\right]^{2}
          = \int d^{3}x \left[\xi_{L}(x)\right]^{2} \,\,.
\end{equation}
Note that these expressions explicitly work at large scales,
where the perturbative expansion is valid.
While the above derivation was for Eulerian bias, Appendix~\ref{sec:d2d_LPT}
shows that the same limit yields
\begin{equation}
{\cal S}(k \rightarrow 0) = b^{L}_{2} Q_{13}(0)
\end{equation}
implying that $b^L_2 = b^E_2$. This is a different relationship that what we 
obtained within the peak-background split, reflecting the different assumptions
made.
 
\begin{figure}
\begin{center}
\leavevmode
\includegraphics[width=3.0in]{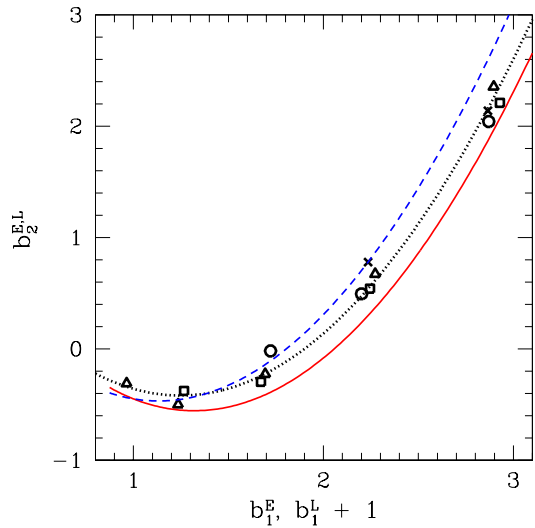}
\end{center}
\caption{Measurements of $b_{2}$ vs. $b_{1}$ for our $c{\rm CDM}$ simulations. The symbols are
as in Fig.~\ref{fig:b1b2}. The solid [red] and dashed [blue] lines are the peak-background
split predictions for the Lagrangian and Eulerian bias, assuming a Sheth-Tormen \cite{STbias} 
mass function. The dotted line is a simple quadratic fit to the data.
}
\label{fig:app_b1b2}
\end{figure}

The procedure for determining $b^{E}_{2}$ from our simulations are straightforward, except for one 
subtlety if determining $\delta_{L}$ from the initial particle data. Since the initial particle positions
are tied to a grid, there is an excess of power at the particle Nyquist frequency, which can alias to 
lower frequencies when computing $\left[\delta_{L}^{2}\right]$. To avoid this, we smooth the initial
density field before squaring; this modifies $Q_{13}(0)$ to
\begin{equation}
Q_{13,S}(0) = \int \frac{d^{3}q}{(2\pi)^{3}} \left[P_{L}(q) W(q)\right]^{2} \,\,,
\end{equation}
where $W(q)$ is the Fourier transform of the smoothing kernel (we adopt a Gaussian
of comoving width $5\,h^{-1}$Mpc). 
We then determine $b_{2}$ by fitting the ${\cal S}(k)$ measurements below $k < 0.05\,h\,{\rm Mpc}^{-1}$
to Eq.~\ref{eq:Sk} where $b^E_1$ is determined from the low $k$ limit of the halo-linear density cross
correlation.

Fig.~\ref{fig:app_b1b2} shows the measured bias parameters for our $c$CDM
simulations, compared with the Eulerian and Lagrangian predictions.
Fig.~\ref{fig:b1b2} demonstrates that the observed
relationship well describes the observed ${\cal B}_1 - {\cal B}_2$ 
correlation, and therefore shifts in the acoustic scale. It is important
to emphasize these constraints on $b_2$ are independent of the acoustic
oscillations. In fact, the $b_2$ constraint depends on the $Q_{13}$ contribution
to the power spectrum, which is irrelevant for BAO.

In principle higher order measures or the observed clustering on small(er)
scales contain information about $\mathcal{B}_i$ for the sample of interest,
and we have shown that improving our ability to model the higher order terms
could bear dividends.
Testing such methods is beyond our scope here, and we defer it to 
future work.
 
\subsection{Shifts, Corrected - A Template for Halos}

We are now in a position to construct a template for the BAO feature traced by
halos. The key ingredient is a calibrated ${\cal B}_1-{\cal B}_2$ relationship;
we assume the Sheth-Tormen form of Eulerian peaks-bias here. Assuming an estimate
of the large-scale bias, this fixes ${\cal B}_2/{\cal B}_1$.
We fit the observed power spectrum to
\begin{eqnarray}
\label{eq:halotemplate}
  P_{\rm w}(k, \alpha) &=& b_1 \left[
  \exp\left(-\frac{k^2\Sigma^2}{2}\right) P_L(k/\alpha) \right. \nonumber \\
  &+& \left. \exp\left(-\frac{k^2\Sigma_1^2}{2}\right)
  \frac{\mathcal{B}_2}{\mathcal{B}_1} P_{22}(k/\alpha) \right]
\end{eqnarray}
where $b_1$ and $\Sigma$ are fit parameters, $\Sigma_1$ is determined
{}from linear perturbation theory (Eq.~\ref{eqn:Sigma}).

We show that this procedure returns (almost) unbiased estimates of the
acoustic scale in Table \ref{tab:shiftsDM}.  These results come from
the halo-mass cross-spectrum, which is significantly better determined
than the halo auto-correlation function.  The results from the auto-correlation
function are consistent with the shift being corrected, but the errors are
too large to allow a meaningful constraint with the simulations we have.
In principle one could obtain even more accurate constraints by modeling
each of the perturbation theory terms separately, but this becomes more
model dependent so we don't pursue this line here.

These results allow us to identify sources of systematic errors in the BAO
measurement and estimate their level.  However, before doing so, we first
extend our results from $c$CDM to $\Lambda$CDM.

\section{Implications for $\Lambda$CDM}
\label{sec:implications}

\begin{figure}
\begin{center}
\resizebox{3in}{!}{\includegraphics{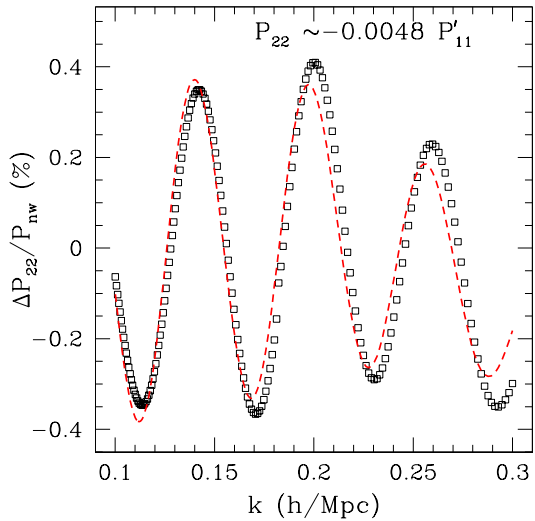}}
\end{center}
\vspace{-0.2in}
\caption{As in Fig.~\ref{fig:derivP00_M004}, except for $\Lambda$CDM.}
\label{fig:derivP00_M000}
\end{figure}

\begin{figure}
\begin{center}
\resizebox{3in}{!}{\includegraphics{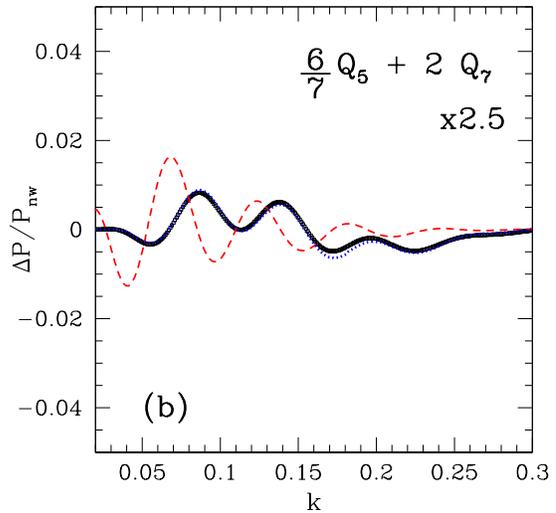}}
\end{center}
\vspace{-0.2in}
\caption{An example of the out-of-phase components, as in
Fig.~\protect\ref{fig:qs4}, except for a $\Lambda$CDM cosmology.
Note the reduced amplitude of the $P_{mn}$ terms, indicating smaller
shifts than for our toy cosmology.  We also note that the factors that
scale $P_{22}$ to the other $Q_n$ combinations appear to be independent
of the underlying cosmology.
}
\label{fig:qs0}
\end{figure}

\begin{figure}
\begin{center}
\leavevmode
\includegraphics[width=3.0in]{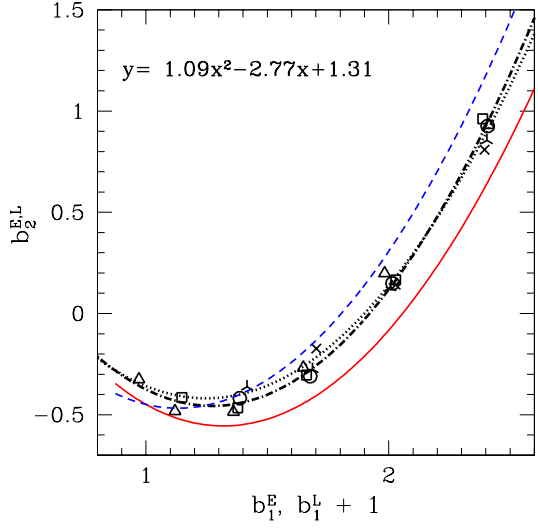}
\end{center}
\caption{As in Fig.~\ref{fig:app_b1b2}, except for $\Lambda$CDM.
These measurements were based on an additional set of simulations employing
$1200^3$ particles in cubic boxes of side $1250\,h^{-1}$Mpc.
The symbols - triangles, squares, circles, stars and crosses correspond to
$z=0$, $0.3$, $0.5$, $0.7$ and $1.0$ respectively.
The dotted line shows the fit from Fig.~\ref{fig:app_b1b2} while the
dot-dashed line is the quadratic fit in the inset.}
\label{fig:M000_app_b1b2}
\end{figure}

In order to extend the results of the previous sections to a $\Lambda$CDM
cosmology, it is useful to summarize the various components of our model
and see how they generalize to a different cosmology. 
\begin{itemize}
\item The shifts in the matter were caused by the $P_{22}$ piece of the 
power spectrum, which well approximated 
a scaled derivative of the linear power spectrum (after subtracting out the broad-band shape). 
The scaling was directly interpretable as the shift in the acoustic scale. 
Fig.~\ref{fig:derivP00_M000} shows that this continues to hold for $\Lambda$CDM, 
with the $z=0$ shift predicted to be $\sim 0.5\%$, in agreement with simulation
results by \cite{Seo08} and analytic arguments by \cite{CroSco08,Tar09}.
\item Extending these results to biased tracers generated additional shifts,
sourced by terms whose oscillatory components resembled scaled versions of 
$P_{22}$. This continues to be true in $\Lambda$CDM with exactly the same
scalings of $P_{22}$; Fig.~\ref{fig:qs0} shows an example. This implies that
the template of Eq.~\ref{eq:halotemplate} as well as the relationship
between the ${\cal B}_m$ and the bias parameters $b_{n}$ continues to hold
for $\Lambda$CDM.
\item The final component was to demonstrate that simple models of halo
bias indeed explained the resulting shifts. The first of these - a peaks-bias
model - is manifestly cosmology independent. The second attempted to empirically
calibrate the bias parameters; Fig.~\ref{fig:M000_app_b1b2} shows a similar
calibration for $\Lambda$CDM. Interestingly, we find a similar cosmology independence
for the empirical calibrations, manifested in both the comparisons with
$c$CDM and with the different redshifts for $\Lambda$CDM.    
\item The above allows us to predict that for $\Lambda$CDM the shift is roughly
an order of magnitude smaller than $c$CDM and is given by 
$\alpha-1 \sim 0.5\% \times D^2\times \mathcal{B}_2/\mathcal{B}_1$.
\end{itemize} 

We can now turn to the systematic error budget for BAO. The expression above
gives the bias in the acoustic scale if one ignored the effects in this paper.
Errors in the calibration of ${\cal B}_2/{\cal B}_1$ directly translate into
an error in the acoustic scale; eg. a $20\%$ error in  ${\cal B}_2/{\cal B}_1$
(approximately how good our toy models are) would correspond to a bias of
$0.1\% \times D^2$.  

Our analysis also demonstrates that the degree of shift is sensitive to the
degree of nonlinearity, or amplitude of the power spectrum.  The out-of-phase
terms scale as one higher power of $P_L$ than the linear terms.  To the
extent that the amplitude is degenerate with a change in bias of the tracer,
uncertainty in the amplitude leads to uncertainty in the shift.  As an example,
if the local slope of the ${\cal B}_2-{\cal B}_1$ relation is $\beta$ and we
imagine holding the large-scale power fixed, a change in the amplitude
$\delta P_L/P_L=\varepsilon$ will induce a change
$\delta{\cal B}_1/{\cal B}_1=-\varepsilon$ and $(2-\beta)\varepsilon$
in the $P_{22}$ term in Eq.~(\ref{eq:Phalo_eff}).
{}From Fig.~\ref{fig:b1b2} we see typical values of $\beta\sim 1$.
Applying the same scaling between $\alpha$ and $P_{22}$ as above this would
lead to a shift in the acoustic scale of $\sim 0.005\,\varepsilon$ for
$\Lambda$CDM.
Thus $10\%$ knowledge of the amplitude of $P_L$ would give $<0.1\%$
uncertainty in $\alpha$.

\begin{figure}
\begin{center}
\resizebox{3in}{!}{\includegraphics{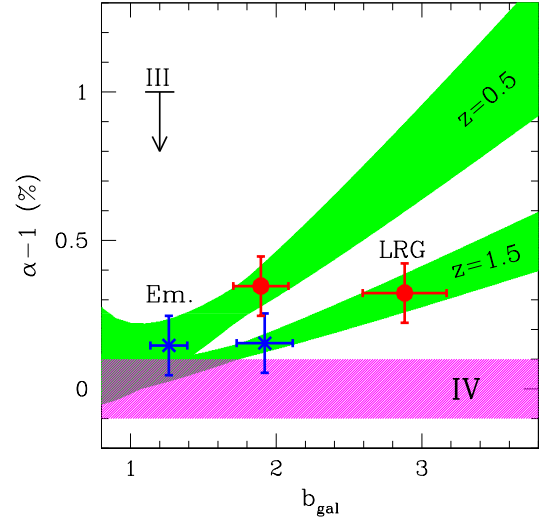}}
\end{center}
\vspace{-0.2in}
\caption{Estimates of the shift (if not corrected) as a function of halo bias
and redshift for our $\Lambda$CDM cosmology. The width of the shaded regions
denotes the error estimated from the difference between the Press-Schechter
and Sheth-Tormen forms in the conversion between ${\cal B}_1$ and
${\cal B}_2/{\cal B}_1$ and demonstrates approximately how errors in the theory
propagate into a residual shift.
We highlight two example populations of halos:
a $b(z=0)=1$ sample [blue crosses] to represent emission line galaxies,
and a $b(z=0)=1.5$ sample [red circles] to represent an elliptical sample.
In both cases, the clustering of the sample is assumed to be constant with
redshift.  The errorbars correspond to a 10\% measurement of the bias.
Also shown are nominal distance accuracies for Stage III (currently underway)
and Stage IV (future) experiments.
}
\label{fig:implications}
\end{figure}

Fig.~\ref{fig:implications} summarizes the systematic error budget, and compares
it to the  observational error goals for Stage III and Stage IV experiments
\cite{DETF}. None of these systematics are expected to be relevant for 
Stage III experiments, and are within a factor of a few of the requirements
for Stage IV experiments.

\section{Galaxies}

The above results have focused on the case of halo samples of a single mass,
but can be generalized to arbitrary combinations of halo samples. 
Of particular interest is the halo model for galaxies (for a review,
see \cite{CooraySheth}), that has been very successful in describing the
large scale clustering of galaxies.

The halo model assumes that all galaxies live in dark matter halos and
the probability of a particular galaxy occupying a given halo depends only
on the halo mass.  Under this assumption the large scale clustering of
galaxies is determined by the clustering of the halos, weighted by the mean
number of galaxies in each halo.
If we denote the mean number of galaxies in a halo of mass $M$ as $N(M)$, the
galaxy power spectrum is
\begin{equation}
\label{eq:Pgal}
  P_{\rm gal} = \frac{\sum_{i,j}^{n} w_i w_j P^{i,j}_{\rm h}}
                     {\left(\sum_i^n w_i\right)^2} \,\,,
\end{equation}
where $P^{i,j}_{\rm h}$ is the halo cross-spectrum for masses $i$ and $j$ and
the weights are determined by the halo mass function, $n_h(M)$, and $N(M)$ via
\begin{equation}
  w_i = n_{\rm h}(M_i) N(M_i) \Delta M_i 
\end{equation} 
Substituting Eq.~\ref{eq:Phalo_eff} for the halo power spectrum and noting that
the exponential damping is (to a good approximation) independent of halo mass,
we find that $P_{\rm gal}$ retains the same structure,
\begin{equation}
\label{eq:galP}
  P_{\rm gal}(k) = \exp\left(-\frac{k^2 \Sigma^{2}}{2}\right)
    \left[{\cal B}_{1,{\rm gal}}P_{L}  + {\cal B}_{2,{\rm gal}}P_{22}\right] .
\end{equation}
but with new coefficients
\begin{eqnarray}
\label{eq:galBn}
  {\mathcal B}_{n, {\rm gal}} &=& \frac{1}{{\mathcal N}^2}
  \int dM_1 n_{\rm h}(M_1) N(M_1) \times \nonumber \\
  & &\int dM_2 \, n_{\rm h}(M_2) N(M_2) {\mathcal B}_{n}(M_1, M_2) \,\,, 
\end{eqnarray}
where the normalizing factor 
\begin{equation}
  {\mathcal N} = \int dM \, n_{\rm h}(M) N(M) \,\,
\end{equation}
is just the mean number of galaxies.
Then, assuming $\mathcal{B}_{n}(M_1,M_2)$ factorizes (or is a sum of 
factorizable pieces), we can simply replace
$b_n$ in our earlier expressions with
\begin{equation}
  b_{n,{\rm gal}} = \frac{1}{\mathcal{N}} \int dM\, n_h(M) N(M) b_n(M)
\end{equation}
Note that, as expected, ${\cal B}_{1, \rm gal}$ simplifies to $b_{\rm gal}^2$
where the galaxy bias $b_{\rm gal}$ is just the weighted sum of the halo
$b_1$'s.  The expressions for cross-spectra of galaxy samples follow a similar
pattern to the halo cross spectra discussed earlier (Eq.~\ref{eq:crossmap}).

\begin{figure}
\begin{center}
\leavevmode
\includegraphics[width=3.0in]{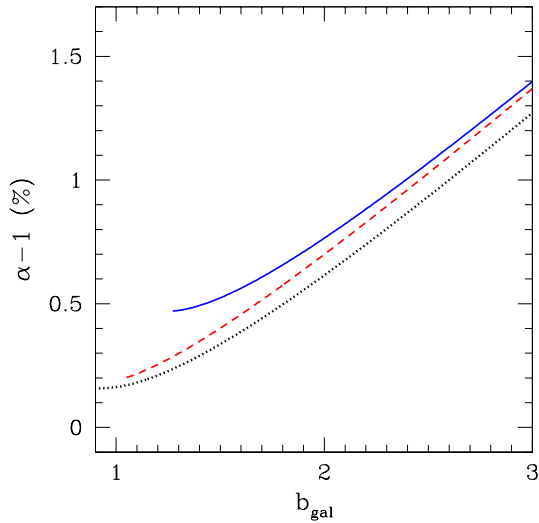}
\end{center}
\caption{The shifts for three example galaxy samples at $z=0$ for $\Lambda$CDM,
as a function of the galaxy bias. The dotted line shows the shifts for halos,
while the dashed [red] and solid [blue] lines are for threshold and satellite 
samples respectively (see the text for more details).
We assume the Eulerian Sheth-Tormen peaks-bias model for definiteness.
}
\label{fig:galaxies}
\end{figure}

For a concrete example, we assume a $N(M)$ of the form,
\begin{equation}
N(M)  = \Theta\left(M-M_{\rm min}\right)\left[1+M/M_1\right] \, .
\end{equation}
We consider two cases : $M_1 = \infty$ or a ``threshold'' sample, and 
$M_1 = 10 M_{\rm min}$ or a ``satellite'' sample; Fig.~\ref{fig:galaxies}
plots the shifts as we vary $M_{\rm min}$ from $10^{12} h^{-1} M_{\odot}$
upwards. 
As one might expected, the increased weighting towards higher halo masses
increases the shifts, although they remain smaller than the required
systematics levels for near future surveys.  The above also suggests that
one could control systematics by re-weighting galaxies, the details of which
will, of course, be population dependent.

\section{Redshift space}

Until now, everything we have done has been in real space, ignoring the
redshift space distortions that arise from peculiar velocities.
The Lagrangian perturbation theory formalism allows us to predict the
effect of large-scale redshift space distortions in a straightforward manner,
although the comparison with simulations is made more difficult and
additional modeling of the observations is required.  We defer a detailed
comparison with simulations to future work, but comment on the trends here.

The power spectrum now becomes anisotropic, $P(k,\mu)$, with $\mu$ the
cosine of the angle between the line-of-sight and $k$ and our previous
results correspond to $\mu=0$.  Including the full $\mu$ dependence
results in different combinations of $Q_n$ entering the expression.  Some
of these are in phase with $P_L$, while some are out of phase.  This
means the shifts in redshift space will be different than those in real
space, as seen in simulations (e.g.~\cite{Seo08}). The in-phase term
($E_{12}$, see Eq.~A73 in \cite{MatsubGal}) is much smaller 
than $P_L$, and we ignore it for now.

If we concentrate on the isotropic piece of the power spectrum \cite{PadWhi08},
we again find that the remaining combinations of $Q_n$ which enter are
proportional to $P_{22}$ and the constants of proportionality appear to
cosmology independent (though they do depend additionally on
$f\equiv d\ln D/d\ln a\simeq\Omega^{0.6}$).
There are no new degrees of freedom introduced theoretically, so in principle
the redshift space shifts are determined from the same modeling as the real
space shifts discussed previously.

For $\Lambda$CDM at $z=0$ the predicted shift in the matter grows from 
$0.5\%$ to $0.75\%$, consistent with the shifts  seen in \cite{Seo08}.
For biased tracers, the effect is to increase the shift by an $f$-dependent
(but roughly bias-independent)
constant. For $f=1$ (corresponding to high $z$), 
this constant is $\sim 0.5\%\ D^2$, while for $f\sim0.5$ (corresponding to $z\sim0$), it is 
$\sim 0.3\%$. If uncorrected, these shifts
could be relevant for future experiments; we leave detailed calibrations to future work.

\section{Discussion}

The propagation of acoustic waves in the early universe provides a
robust means for determining the expansion history of the Universe,
and contraining cosmology.  The standard ruler is calibrated in the
linear regime by observations of the CMB, but observed today with
biased, nonlinear tracers.  Since the acoustic scale is so large,
the effects of bias and nonlinearity on the acoustic scale is small,
but future experiments may have enough sensitivity that it needs to
be taken into account.  We have begun this program here.

Using a set of N-body simulations of an extreme cosmology, in which
the acoustic scale is relatively small and the nonlinearity quite
pronounced, we have shown that shifts in the scale grow quadratically
with the amplitude of the linear theory power spectrum for both the
mass and for dark matter halos.  Motivated by this, and guided by
arguments from Eulerian and Lagrangian perturbation theory, we found
the dominant second-order contribution to the peak shift.
This contribution, $P_{22}$, quite well approximates the derivative of
the acoustic signal, explaining why it leads to a peak shift and allowing
us to estimate how the amplitude of the $P_{22}$ term can be translated into
a shift in the fitted acoustic scale.

For dark matter halos the contribution depends on two bias parameters,
$b_1$ and $b_2$, allowing in principle arbitrary shifts of the acoustic
scale.  We showed however that dark matter halos, which will be the hosts
of any galaxies we observe, obey a relation between $b_1$ and $b_2$ which
is relatively well predicted by the peak-background split.
Once these two terms are related the acoustic scale for nonlinear, biased
tracers can be accurately determined, allowing high fidelity measurements of
distances using baryon acoustic oscillations.

We have described how redshift space distortions affect the scale shifts
within the context of Lagrangian perturbation theory, where they increase
the shift by tens of percent at low $z$ and about a factor of $2$ at high $z$.

In this paper we have concentrated on the effects of nonlinear gravitational
evolution and halo biasing, showing that these effects could be understood
at the $0.1\%$ level.  At this level of precision the inclusion of additional
physics, such as differential evolution of the baryonic and dark matter
components or the details of galaxy formation, may enter and investigations
in this direction should be undertaken.  We have also assumed that the linear
theory template is perfectly understood, which also deserves further study.
Such investigations must be part of any future high-precision BAO experiment.

The simulations presented in this paper were carried out using computing
resources of the National Energy Research Scientific Computing Center and
the Laboratory Research Computing project at Lawrence Berkeley National
Laboratory.
NP is supported by NASA HST-HF-01200.01 and LBNL.
MW is supported by NASA and the DoE.
This research was additionally supported by the Laboratory Directed Research
and Development program at Lawrence Berkeley National Laboratory, and by
the Director, Office of Science, of the U.S.  Department of Energy under
Contract No. DE-AC02-05CH11231.

\appendix

\section{Expressions for $\mathbf{P_{mn}}$ for $\mathbf{m=2,n=2}$}
\label{sec:qdefs}

The expressions for $P_{mn}$ for a biased tracer (in both Eulerian and
Lagrangian perturbation theory) can be simply written by defining
\cite{MatsubGal}
\begin{equation}
\label{eq:qndef}
  Q_{n}(k) = \frac{k^{3}}{4\pi^2} \int_{0}^{\infty}dr\,P_{L}(kr) 
  \int_{-1}^{1}dx\,P_{L}(k\sqrt{y}) \tQ_{n}(r,x) \,\,,
\end{equation}
where $y(r,x) = 1 + r^{2} - 2rx$, $P_L$ is the linear power spectrum and
the $\tQ_{n}$ are given by
$$\setlength\arraycolsep{0.1em}
\begin{array}{ll}
\displaystyle \tQ_{1} = \frac{r^2(1-x^2)^2}{y^2}, &\displaystyle  \tQ_{2} = \frac{(1-x^2)rx(1-rx)}{y^2}, \\
\displaystyle \tQ_{3} = \frac{x^2(1-rx)^2}{y^2}, &\displaystyle   \tQ_{4} = \frac{1-x^2}{y^2}, \\
\displaystyle \tQ_{5} = \frac{rx(1-x^2)}{y}, &\displaystyle   \tQ_{6} = \frac{(1-3rx)(1-x^2)}{y}, \\
\displaystyle \tQ_{7} = \frac{x^2 (1-rx)}{y}, &\displaystyle   \tQ_{8} = \frac{r^2(1-x^2)}{y}, \\
\displaystyle \tQ_{9} = \frac{rx(1-rx)}{y}, &\displaystyle  \tQ_{10} = 1-x^2, \\
\multicolumn{2}{l}{\displaystyle \tQ_{11} = x^2,\,\, \displaystyle  \tQ_{12}=rx,\,\, \displaystyle \tQ_{13}=r^2} 
\end{array}
$$    

\section{Computing $\mathbf{\langle\delta_L^n\delta_{\rm \bf obj}\rangle}$
in resummed LPT}
\label{sec:d2d_LPT}

We present expressions for evaluating
$\langle [\delta_L^{n}]\delta_{\rm obj} \rangle$
in resummed Lagrangian perturbation theory, where $\delta_{\rm obj}$ is the
density field of biased tracers.  These expressions follow \cite{MatsubGal}
and we refer the reader there for detailed calculations.

The density field for a biased tracer can be defined by the displacement field
$\mathbf{\Psi}(\mathbf{q})$ and a function of the smoothed initial density
field in Lagrangian space, $F[\delta_{L}(\mathbf{q})]$, as
\begin{equation}
\delta_{\rm obj}({\mathbf x}) = \int d^{3}q F[\delta_{L}(\mathbf{q})] 
   \delta^{(3)}_{D}(\mathbf{x} - \mathbf{q} - \mathbf{\Psi}) \,\,,
\end{equation}
where $\mathbf{x}$ and $\mathbf{q}$ are the Eulerian and Lagrangian positions
and $\delta^{(3)}_D$ is the 3D Dirac $\delta$ function.
We implicitly assume that the argument to $F$ has been smoothed on some scale
much smaller than the large scales of relevance here, allowing us to ignore
the smoothing here (see \cite{MatsubGal} for a detailed justification).
We cross correlate this with a field defined by $\exp(i\lambda \delta_{L})$;
$\delta^{n}_L$ is then simply obtained by taking the $n$-th derivative with
respect to $\lambda$ and setting $\lambda$ to zero.

The cross-power spectrum of the two fields is then given by
(compare to Eq.~9 of \cite{MatsubGal})
\begin{eqnarray}
  {\cal H}(k) = \int d^{3}q e^{-i \mathbf{k} \mathbf{q}}
  \left[ \int_{-\infty}^{\infty}
  \frac{d\lambda_{2}}{2\pi} \widetilde{F}(\lambda_2) \times \right.\nonumber \\
  \left. \left\langle e^{i\left(\lambda_1 \delta_L(\bq_1) +
  \lambda_2 \delta_L(\bq_2)\right) +
  i \mathbf{k}\mathbf{\Psi}(\bq_2)}\right\rangle \right] \,\,,
\label{eqn:Hfirstdef}
\end{eqnarray}
where $\bq = \bq_1 - \bq_2$ and $\widetilde{F}$ is the Fourier transform of
$F$.  The correlators of interest are then given by
\begin{equation}
  \langle [\delta_L^{n}] \delta_{\rm obj} \rangle = 
  \frac{1}{i^{n}} \frac{d^{n} {\cal H}}{d\lambda_{1}^{n}} \bigg|_{\lambda_1=0} \,\,.
\label{eq:momdef}
\end{equation}
The algebra now follows through as in \cite{MatsubGal} using the cumulant
expansion theorem, and collecting all zero-lag correlators to yield
(compare to Eq.~24 in \cite{MatsubGal}),
\begin{widetext}
\begin{eqnarray}
{\cal H}(k) & = &  \exp\left[\sum_{m=1}^{\infty} \frac{(-1)^{m}}{(2m)!} B^{0\,0}_{0\,2m}(\bk,\bq)\right] \int 
d^{3}q e^{-i\bk\bq} \int_{-\infty}^{\infty} \frac{d\lambda_{2}}{2\pi} \widetilde{F}(\lambda_2) \times \nonumber \\
& & e^{-\lambda_1^2 \sigma^2/2 - \lambda_2^2 \sigma^2/2} 
\exp\left[-\lambda_1 \lambda_2 \xi(|\bq|) + \sum_{n_1+n_2 \ge 1}^{\infty} \sum_{m_2 \ge 1}^{\infty}
\frac{i^{n_1 + n_2 + m_2}}{n_1! n_2! m_2!}
\lambda_1^{n_1} \lambda_2^{n_2} B^{n_1 n_2}_{0\,m_2}(\bk,\bq)\right]
\label{eq:hdef}
\end{eqnarray}
\end{widetext}
where 
\begin{equation}
\xi(|\bq|) = \langle \delta_{L}(\bq_1) \delta_{L}(\bq_2)\rangle; \,\, \sigma^2 = \xi(0) \,\,,
\end{equation}
and 
\begin{equation}
  B^{n_1 n_2}_{0\,m_2} \equiv
  \langle [\delta_{L}(\bq_1)]^{n_1} [\delta_{L}(\bq_1)]^{n_2}
  [\mathbf{k}\mathbf{\Psi}(\bq_2)]^{m_2} \rangle_c \,\,,
\end{equation}
with $\langle \cdots \rangle_c$ denoting the connected moments.

Given Eq.~\ref{eq:hdef}, it is straightforward (if tedious) to compute 
expressions for the correlators in Eq.~\ref{eq:momdef}.
Of particular interest to us here is the $k\rightarrow 0$ limit of
$\langle [\delta_L^{2}] \delta_{\rm obj} \rangle$ which, as in Eulerian
perturbation theory, is given by
\begin{equation}
  {\cal S}(k \rightarrow 0) = b^L_{2} Q_{13}(0) \,\,, 
\label{eqn:Skto0}
\end{equation}
with $b^L_2$ defined by 
\begin{eqnarray}
b^L_n & \equiv & \int_{-\infty}^{\infty} \frac{d\lambda}{2\pi} 
    e^{-\lambda^{2}\sigma^2/2}\widetilde{F}(\lambda) (i\lambda)^{n}  \nonumber \\
&=& \frac{1}{\sqrt{2\pi} \sigma} \int_{-\infty}^{\infty} d\delta e^{-\delta^2/2\sigma^2} \frac{d^{n}F}{d\delta^n}
\end{eqnarray} 
While the full formalism is required in general, we note that this $k\to 0$
limit can be obtained more simply by dropping the $\bk\cdot\mathbf{\Psi}$
term in Eq.~\ref{eqn:Hfirstdef}, leaving only Gaussian fields in the exponent.
These fields have only a second connected moment, thus only the
$\lambda_1\lambda_2\xi(|\bq|)$ term survives in the last exponential of
Eq.~\ref{eq:hdef} and expanding this exponential gives Eq.~\ref{eqn:Skto0}.

\onecolumngrid

\bibliography{baoshift}
\bibliographystyle{apsrev}

\end{document}